\documentclass[12pt, draftclsnofoot, onecolumn]{IEEEtran}
\renewcommand\IEEEkeywordsname{Index Terms}
\usepackage{blindtext, graphicx}
\usepackage{csquotes}
\usepackage{cite}
\ifCLASSINFOpdf
\else
\fi

\usepackage[cmex10]{amsmath}
\usepackage{optidef}
\usepackage{array}
\usepackage[tight,footnotesize]{subfigure}
\usepackage{stfloats}
\usepackage{float}
\def\*#1{\mathbf{#1}}

\usepackage{color}
\definecolor{light-gray}{gray}{0.5}

\usepackage{amsfonts}

\begin{document}
\title{How URLLC can Benefit from \\ NOMA-based Retransmissions}

\author{Rados\l{}aw~Kotaba,
        Carles~Navarro~Manch\'{o}n,
        Tommaso~Balercia,
        and~Petar~Popovski,~\IEEEmembership{Fellow,~IEEE}%
        
\thanks{Manuscript received February 17, 2020. The work of R. Kotaba and P. Popovski was funded by the European Research Council (ERC), Consolidator Grant Nr. 648382 (WILLOW), under the European Union Horizon 2020 research and innovation program.
The work of R. Kotaba and C. N. Manch\'{o}n was supported by the Virtuoso project funded by Innovation Fund Denmark.
}
\thanks{R. Kotaba, C. N. Manch\'{o}n and P. Popovski are with the Department of Electronic systems, Aalborg University, Aalborg 9220, Denmark  (e-mail:
rak@es.aau.dk; cnm@es.aau.dk; petarp@es.aau.dk).}
\thanks{T. Balercia is with Nvidia Corporation, Santa Clara 95051, California.}
\thanks{R. Kotaba and T. Balercia were with Intel Mobile Communications, Aalborg 9220, Denmark.}
}

\maketitle

\begin{abstract}
Among the new types of connectivity unleashed by the emerging 5G wireless systems, Ultra-Reliable Low Latency Communication (URLLC) is perhaps the most innovative, yet challenging one. 
Ultra-reliability requires high levels of diversity, however, the reactive approach based on packet retransmission in HARQ protocols should be applied carefully to conform to the stringent latency constraints.
The main premise of this paper is that the NOMA principle can be used to achieve highly efficient retransmissions by allowing concurrent use of wireless resources in the uplink.
We introduce a comprehensive solution that accommodates multiple intermittently active users, each with its own HARQ process.
The performance is investigated under two different assumptions about the Channel State Information (CSI) availability: statistical and instantaneous.
The results show that NOMA can indeed lead to highly efficient system operation compared to the case in which all HARQ processes are run orthogonally.
\end{abstract}

\IEEEpeerreviewmaketitle

\section{Introduction}\label{sec:intro}

The fifth generation (5G) wireless networks are slowly becoming a reality. 
While historically the primary motivation behind each new generation was to increase data rates, coverage and other metrics related to the quality of experience of the users, 5G promises to be more than just an incremental improvement over previous technologies \cite{3gpp:definiton,urllc:Popovski}.
This shift is driven by a growing popularity and rapid advancements in the area of Internet of Things (IoT) which represents a different, non-human-centric communication paradigm.
Among those new, emerging applications, especially prominent are those that fall into the category of ultra-reliable low-latency communications (URLLC).
Examples of such use cases include: smart cities, factory automation (Industry 4.0)\cite{urllc:factory}, and tactile Internet (involving remote motion control, telesurgery, etc.)\cite{urllc:telesurgery}.
To enable those demanding applications, the underlying network will need to provide MAC-layer end-to-end latencies from $0.5$ to few milliseconds and reliability (defined as the probability of successful delivery of the packet within the stipulated latency) above $99,999\%$ \cite{3gpp:req}.

Designing an efficient URLLC system capable of meeting the aforementioned requirements poses a significant challenge, especially considering the fundamental tradeoffs between latency, reliability, spectral efficiency, and power consumption \cite{tradeoffs}.
While it has been shown that on their own legacy systems are either not able to operate in URLLC regime \cite{lte1}, or become prohibitively inefficient \cite{not:enough}, many of the concepts they use are still valid and can be adapted to this new paradigm. Diversity-providing mechanisms are particularly crucial, since they are an unavoidable necessity when facing stringent reliability requirements.

One such mechanism is hybrid automatic repeat request (HARQ), which provides diversity in a reactive way upon reporting of an error by the receiver.
Its flexibility and the potential to offer significant gains have been thoroughly studied both theoretically~\cite{harq:theory} and in practical scenarios~\cite{harq:hspa} which led to the implementation of HARQ in the third generation system HSPA and onwards. While applying HARQ in URLLC is challenging due to the stringent latency constraint, we note that the alternative for reaching high reliability through one-shot transmission~\cite{oneshot} is very inefficient in terms of power and, whenever feasible, some form of HARQ is highly desirable.
As shown in~\cite{nurul:timing}, even with latency  budget as low as $1$ ms, the new 5G features including: shortened transmission time intervals (TTIs), higher subcarrier spacing and improved processing times will allow for at least one retransmission opportunity.

While generally beneficial, especially as a mechanism to enhance reliability, HARQ in URLLC should be designed in a lean way and avoid inefficiencies. First, as the amount of time-frequency resources in the system is finite, the need to accommodate both new packets and retransmissions increases the probability of queuing which is especially detrimental for URLLC. Second, as the system preserves the previous unsuccessful copies of the packet, retransmission of the full payload can be wasteful. 
Meanwhile, practial systems prefer to work with fixed-size resources where adapting the size of the retransmissions is not possible.

The shortcomings of HARQ can be mitigated with the help of non-orthogonal multiple access (NOMA). This technique involves transmitting multiple packets over the same time-frequency resources thereby intentionally introducing interference. 
Due to its ability to accommodate more users and reduce latency, NOMA has been identified by researchers as one of the enablers of URLLC \cite{urllc:Poor}
. For a comprehensive overview of this topic and a discussion on different existing variants of NOMA reader is directed to \cite{noma:general1}\cite{noma:general2}. 
Our motivation for using NOMA is the fact that it can address the HARQ inefficiencies and allow efficient use of the time-frequency resources.

\subsection{Related work}\label{sec:intro:prior}
On their own, both HARQ and NOMA topics have been extensively covered in the literature. 
In \cite{CC:larsson}, the authors optimize the average power of HARQ with finite number of retransmissions and a given outage probability target. The Chase combining (CC) variant is assumed, Rayleigh fading channel and a single bit feedback.
The incremental redundancy (IR) type HARQ is studied in \cite{IR:szczecinski}, where the aim is to maximize the throughput for a given reliability constraint. This is achieved through rate adaptation, however the assumption of a full buffer used there might not be suitable for all URLLC applications.
In \cite{avranas:energy} and \cite{avranas:throughput}, the authors investigated HARQ explicitly in the URLLC context by considering transmission of short packets (finite blocklength regime) over AWGN channel. Moreover, in their optimization problems authors consider the impact of the feedback delay and overall energy budget.

While the literature on uplink NOMA is not as extensive as on its downlink counterpart, some interesting contributions can be found in \cite{noma:uplink1}, \cite{noma:journal1}. The former provides insights into the achievable sum-rate and outage probability with a given transmit power, while the latter discusses rate and power allocation scheme that ensures required probability of error. As far as solutions combining both HARQ and NOMA are concerned, the literature is even more scarce. Some of the reported works include \cite{noma-harq-ir,noma-harq-cc,choi:nomaharq}, but except for the last one, they do not consider uplink scenario which entails radically different system model. To the best of the authors knowledge, none of the contributions on NOMA and HARQ 
deal with the comprehensive, multi-user scenario where the amount of resources is finite and the effects of queuing are considered.

\subsection{Contributions}\label{sec:intro:contributions}
In this work we investigate the performance of the uplink OMA and NOMA systems employing HARQ mechanism. 
Considered framework involves very limited number of retransmission opportunities and tight reliability constraints which are meant to conform to the URLLC use case and hence provide useful insights into the design of practical systems.
As a main contribution of this paper, we develop a comprehensive solution involving power allocation and packet scheduling that efficiently accommodates multiple intermittently active URLLC users, each with its own HARQ process, over a finite pool of resources.
We achieve this by decoupling the two problems. First, we formulate the power allocation problem as a minimization of the average transmit power subject to the reliability constraints. This is done by finding optimal error targets per each HARQ round.
Next, a joint scheduling problem is considered, where we develop a simple heuristic that allows to make a decision which packets should be prioritized in case of insufficient resources based on the optimal error targets and transmit powers determined in the earlier step.

The solution outlined above is developed in two variants, based on OMA and NOMA principle. Furthermore, for each of them we propose two different approaches depending on the type of CSI available: \textit{Statistical CSI}, where only the distribution of channel realizations is known and \textit{Instantaneous CSI}, where additionaly the channel conditions of the next transmission (and only the next) are known.
In the former case, we study and compare the performance of CC and IR HARQ techniques assuming asymptotic (infinite) blocklength. In the Instantaneous CSI case where the channel at hand becomes AWGN, we develop the methodology and analyze OMA- and NOMA-HARQ in incremental redundancy mode under the finite-blocklength assumption.
The two CSI scenarios are meant to provide bounds on how the channel knowledge can impact the performance.
The proposed approaches are evaluated by means of Monte Carlo simulations, revealing that our NOMA schemes can effectively deal with more than twice as much URLLC load as their OMA counterparts using the same amount of channel resources. This increase in system capacity comes at the expense of only a slight increase in transmit power.

This contribution extends the prior study \cite{kotaba:icc} by introducing significant changes.
Most notably, as the activation of users is no longer deterministic and the total user population is larger than the amount of channel resources queuing issues need to be taken into account.
The signal model now includes the effect of distant-dependent large scale fading, which impacts the power assignment in NOMA, as some UEs become preferable to the others. Furthermore, unlike in \cite{kotaba:icc}, we do not restrict the pairing in NOMA to be only between new packets and retransmissions. Instead, we generalize the approach and allow the packets to be scheduled non-orthogonally in whichever way that minimizes the total power spent. 

	\indent The rest of the paper is organized as follows. 
	In Section \ref{sec:model} we describe the system and signal model.
	In Section \ref{sec:fading} we go into the details of optimal error targets and power allocation for OMA and NOMA with statistical CSI.
	In Section \ref{sec:finite} we extend the discussion to the known channel case and finite blocklength communication.
	In Section \ref{sec:scheduling} we discuss briefly the scheduling and resource allocation technique.
	In Section \ref{sec:res} we	present the simulation results together with their thorough discussion.
	Lastly, in Section \ref{sec:concl} we offer final conclusions that close the paper.

\section{System model}\label{sec:model}

In this work, we consider a single cell serving the uplink traffic of $N$ devices running URLLC applications. 
We assume that the packets of each UE are of the same, fixed size and span $K$ channel uses, i.e. symbols. Moreover, they carry the same amount of $B$ information bits leading to equal rates which we denote as $R=B/K$ [bits/symbol].
The $K$ channel uses that constitute a packet occupy a contiguous block of time-frequency resources which we will interchangeably refer to as TF-block or slot.
A single TF-block is considered smaller than the coherence bandwidth/time and distinct slots experience independent Rayleigh fading.
The number of available TF-blocks is limited to $W$ per uplink phase, and the base station's (BS) goal is to best distribute them between UEs' transmissions.
The generation of new packets at each device $j$ is intermittent and occurs with probability $b$. Whenever new packet appears, UE sends a scheduling request (SR) for that packet in the next available uplink phase and consequently receives the grant from the BS with instructions regarding time-frequency resource allocation and appropriate transmit power. It is further assumed that this step is error-free and happens in parallel with the usual exchange of packets that are carrying payload, i.e. there are dedicated resources for SRs and in a single uplink phase UE can send both the previously scheduled packets as well as a new request. These assumptions are reasonable in the URLLC context considering the stringent latency requirements. As such, we can view the scheduling handshake procedure as transparent as it simply creates a constant offset between the arrival of the new packet at UE's buffer and the moment it is transmitted. Hence, in the remainder of this paper we will simply say that in each uplink phase device will transmit a new packet with probability $b$.

Due to the latency requirement of URLLC, we assume that once a new packet is generated it can be transmitted only during the next $L+1$ uplink phases and is dropped otherwise. 
Consequently, unsuccessful packets can be retransmitted during that window to increase the reliability (up to $L$ times if every opportunity is used).
Two variants of the HARQ mechanism are considered for this: Chase combining (CC) and incremental redundancy (IR).\\
\indent Following the NOMA principle, in this work we admit the possibility of users sharing the same resources. Let us denote by $\mathcal{I}^i$ the set of indices of the UEs transmitting over $i$-th TF-block. The complex baseband signal received over its $K$ channel uses can be written as
\begin{equation}
\*y_i = \sum_{j \in \mathcal{I}^i}\sqrt{P_{i,j}}g_{i,j}\*x_{i,j} + \*n_i
\label{eq:sig_mod}
\end{equation}
where $P_{i,j}\in \mathbb{R}$ is the transmit power of user $j$ in TF-block $i$, $g_{i,j}\in \mathbb{C}$ is the channel between $j$-th UE and the BS over the $i$-th TF-block, $\*x_{i,j}\in \mathbb{C}^K$ are the complex transmitted symbols assumed to be Gaussian distributed with zero mean and unit variance and $\*n_i\in \mathbb{C}^K$ is complex additive white Gaussian noise with zero mean and variance $\sigma^2$. 
The channel coefficients are given by $g_{i,j}=\frac{h_{i,j}}{\sqrt{d_j^{\alpha}}}$, where $h_{i,j}$ is the Rayleigh fading component, which is independent and identically distributed (i.i.d) zero mean circularly symmetric complex Gaussian (ZMCSCG) random variable with unit variance, while $d_j^{\alpha}$ is a pathloss term accounting for the distance between UE $j$ and the BS. 
The distance itself is uniformly distributed as $d_j \sim U(D_{min},D_{max})$. The realizations of $h_{i,j}$ change between different transmissions while the distance $d_j$ remains constant for a particular user.\looseness=-1

In this work we consider two different scenarios that provide the bounds on the performance of presented HARQ schemes.
\subsubsection{Statistical CSI}\label{sec:model:stat}
Similarly to the work presented in \cite{kotaba:icc} we assume here that at the time of scheduling new transmissions the base station has only a statistical knowledge about the future channel realizations $h_{i,j}$, i.e. that they are i.i.d. ZMCSCG. BS knows however the distances $\*d$ of all users and hence knows the variance of $\*g$.
\subsubsection{Instantaneous CSI}\label{sec:model:known}
In this scenario we assume that BS knows the CSI of the next transmission at the time of performing the scheduling, i.e. it knows the channel coefficients $\*g$ for the immediate uplink stage, but not the ones coming afterwards\footnote{In practice, obtaining the CSI involves auxiliary procedures that can deteriorate the reliability. By neglecting these we gain an insight into the upper bound of the performance in such scenario.}.

\subsection{Base Station operation}\label{sec:model:recv}

\begin{figure}[t!]%
\centering
\includegraphics[width=0.53\linewidth]{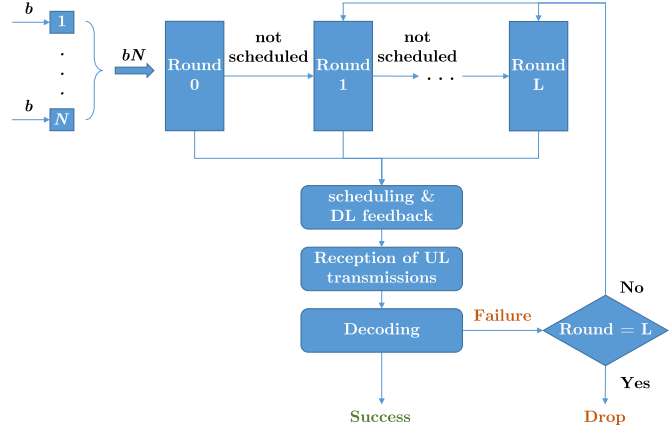}
  \caption{Receiver operation}
  \label{fig:sys_model}
\end{figure}
A simplified diagram explaining the principle of operation of the receiver is shown in Fig.~\ref{fig:sys_model}. 
As discussed earlier, UEs generate new packets independently with probability $b$, so the resulting total number of scheduling requests received by the BS is given by a binomial distribution with $N$ trials and success probability $b$. 
The new packets are referred to as being in state/round 0, while all those that arrived earlier and failed the decoding (or were not transmitted) belong to any of the other $1,\dots,L$ rounds. 
The packets from all rounds are jointly scheduled by the BS, which determines the appropriate assignment of TF-blocks and power levels. The exact procedure governing this step is described in detail in section \ref{sec:scheduling}. The information regarding scheduling and power allocation is then signaled to all concerned UEs so that they can perform coordinated transmission in the upcoming uplink phase. Note that, when the number of resources $W$ is finite, it might not be possible to schedule all the packets, in which case they are moved directly to their next round as if they failed or were transmitted with power $0$. In the decoding step, if NOMA is employed, it is assumed that the receiver is capable of SIC and depending on the use case we will consider either optimal or fixed decoding order.

\section{HARQ with Statistical CSI }\label{sec:fading}

\subsection{OMA-HARQ}\label{sec:fading:oma}
Let us start by analyzing a simpler approach where the base station is allowed to schedule uplink transmissions only in an orthogonal manner, dedicating one TF-block for each packet.
The SNR of the packet received from user $j$, conditioned on its power $P_j$ and distance from the BS $d_j$, is distributed exponentially according to the pdf
\begin{equation}
    f_e \left( x; \frac{P_j}{d_j^{\alpha}\sigma^2} \right) = \frac{d_j^{\alpha}\sigma^2}{P_j} e^{-\frac{x d_j^{\alpha}\sigma^2}{P_j}}.
    \label{eq:snr_dist}
\end{equation}
Taking into account prior unsuccessful transmissions ans assuming CC is used, the decoding failure probability after $l$-th attempt (counting from $0$ as the initial one) is given by \cite{harq:theory}
\begin{equation}
p_{er,cc_j}^{(l)} = \Pr \Bigg\lbrace \log_2 \left( 1+\sum_{i=0}^{l} \mathrm{SNR}_j^{(i)} \right) < R \Bigg\rbrace
\end{equation}
where $\mathrm{SNR}_j^{(i)}$ is the SNR of $j$-th UE's packet in its $i$-th attempt. When IR-HARQ is used, then
\begin{equation}
p_{er,ir_j}^{(l)} = \Pr \Bigg\lbrace \sum_{i=0}^{l} \log_2 \left(1+\mathrm{SNR}_j^{(i)} \right) < R \Bigg\rbrace.
\end{equation}
The two expressions can be rearranged to depend only on the last packet realization since all the previous SNRs are already known
\begin{equation}
p_{er,cc_j}^{(l)} = \Pr \Bigg\lbrace \mathrm{SNR}_j^{(l)} < 2^R - 1 - \sum_{i=0}^{l-1}\mathrm{SNR}_j^{(i)} = \gamma_{cc_j}^{(l)} \Bigg\rbrace
\label{eq:errB}
\end{equation}
\begin{equation}
p_{er,ir_j}^{(l)} = \Pr \Bigg\lbrace \mathrm{SNR}_j^{(l)} < \frac{2^R}{\prod_{i=0}^{l-1}\left(1+\mathrm{SNR}_j^{(i)}\right)} - 1  = \gamma_{ir_j}^{(l)}\Bigg\rbrace.
\label{eq:errBIR}
\end{equation}
To simplify the notation we introduce the terms $\gamma_{cc_j}^{(l)}$ and $\gamma_{ir_j}^{(l)}$ denoting a ``residual SNR'' which is the amount of signal power needed until the packet can be decoded (at $l=0$ simply equal to  $2^R-1$).
Throughout this paper we will typically omit the ${}_{cc} / {}_{ir}$ subscript since each method is discussed in a dedicated section making it clear which definition is used.

By combining eq. \eqref{eq:snr_dist} with either of the two \eqref{eq:errB}, \eqref{eq:errBIR} the error probability is obtained:
\begin{equation}
p_{er_j}^{(l)} = 1 - e^{-\frac{\gamma_{j}^{(l)}d_j^{\alpha}\sigma^2}{P_j^{(l)}}}.
\label{eq:err1U}
\end{equation}
It further follows from \eqref{eq:err1U} that the minimum power required to achieve certain target error $p_{er_j}^{(l)} = \epsilon_j^{(l)}$ is
\begin{equation}
P_j^{(l)} = -\frac{\gamma_{j}^{(l)}d_j^{\alpha}\sigma^2}{\ln(1-\epsilon_j^{(l)})}.
\label{eq:P1U}
\end{equation}
Because the BS's goal is to spend (on average) as little power on a packet as possible while providing certain reliability guarantees, we define the following optimization problem
\begin{mini!}|s|[2]           	
    {\epsilon_j^{(l)}}           	
    {P_j^{(l)} + \int_{0}^{\gamma_{j}^{(l)}} f_e \left( x_l; \frac{P_j^{(l)}}{d_j^{\alpha}\sigma^2} \right) \Psi_j^{(l+1,L)} \left(\gamma_j^{(l+1)},\frac{\Theta_j^{(l)}}{\epsilon_j^{(l)}}\right) dx_l \label{eq:globObj1}}
    {\label{eq:optGlob}}   	
    {\Psi_j^{(l,L)} (\gamma_j^{(l)},\Theta_j^{(l)}) =} 
    \addConstraint{\prod_{i=l}^{L}\epsilon_j^{(i)}}{\leq \Theta_j^{(l)} \label{eq:globCon1}}  
\end{mini!}
where $\Theta_j^{(l)}=\frac{\epsilon_{tar}}{\prod_{i=0}^{l-1}\epsilon_j^{(i)}}$ is the remaining error budget resulting from the previous transmission attempts and the overall target $\epsilon_{tar}$ (such as $10^{-5}$ in URLLC). The problem \eqref{eq:optGlob} can be summarized as follows.
For a given packet, currently at round $l$, BS needs to decide on its next error target $\epsilon_j^{(l)}$ that will minimize the expected power moving forwards. The objective (cost) is composed of two terms. First, the power $P_j^{(l)}$ spent in the immediate round, which is directly related to the chosen error target via (8). Second, the expected additional power that will be spent if the packet fails.
Note that $\epsilon_j^{(l)}$ impacts the second part in two ways: it determines the remaining error budget $\Theta_j^{(l+1)}$ and, through $P_j^{(l)}$, the distribution of the SNR of the current transmission $x_l$ that affects the new residual SNR $\gamma_j^{(l+1)}$. 
Depending on the HARQ mode, the relationship between $x_l$ and $\gamma_j^{(l+1)}$ is captured by either \eqref{eq:errB} or \eqref{eq:errBIR}.
In general, the problem \eqref{eq:optGlob} is difficult as it involves the recursive term $\Psi_j^{(l+1,L)}$ which contains $\Psi_j^{(l+2,L)} $, etc., that all require finding an optimal target.
We will now consider the two special cases that arise when CC or IR is used.

\subsubsection{Chase Combining}\label{sec:fading:oma:cc}
In case of CC mode of HARQ, the optimization problem is greatly simplified, which is captured by the following theorem:
\newtheorem{theorem}{Theorem}
\begin{theorem}
When using Chase Combining, the individual, per-stage error targets that minimize the expected power depend only on the remaining error budget $\Theta_j^{(l)}$. 
Due to the lack of dependency on other parameters, in particular rate $R$ and residual SNRs, the recursive problem \eqref{eq:optGlob} becomes equivalent to
\begin{mini!}|s|[2]           	
    {\epsilon_j^{(l)},\hdots,\epsilon_j^{(L)}}           	
    {\left(-\gamma_{j}^{(l)}d_j^{\alpha}\sigma^2 \right)
\sum_{i=l}^{L} \frac{1}{\ln(1-\epsilon_j^{(i)})}\prod_{k=l}^{i-1}\frac{\ln(1-\epsilon_j^{(k)}) + \epsilon_j^{(k)}}{\ln(1-\epsilon_j^{(k)})} \label{eq:optSimpleObj}}
    {\label{eq:optSimple}}   	
    {\Psi_j^{(l,L)} (\gamma_j^{(l)},\Theta_j^{(l)}) =} 
    \addConstraint{\prod_{i=l}^{L}\epsilon_j^{(i)}}{= \Theta_j^{(l)} \label{eq:optSimpleCon}}  
\end{mini!}
\label{theorem:1}
\end{theorem}
The proof of Theorem \ref{theorem:1} can be found in Appendix \ref{sec:appendix:cc_proof}.
By minimizing $\Psi_j^{(0,L)} (2^R-1,\epsilon_{tar})$ using the definition \eqref{eq:optSimple} stated in the Theorem \ref{theorem:1}, the BS can determine all targets $\boldsymbol{\epsilon} = \left[ \epsilon^{(0)},\ldots,\epsilon^{(L)} \right]$ in advance. While the analytical approach is not tractable, numerical solutions can be obtained rather easily.
Moreover, since the final target error rate $\epsilon_{tar}$ and the maximum number of retransmissions $L$ are typically system-wide parameters with limited number of configurations, the sequence $\boldsymbol{\epsilon}$ do not require frequent updates and is identical for all UEs.

\subsubsection{Incremental Redundancy}\label{sec:fading:oma:ir} 
When the IR-type HARQ is used, determining optimal error targets is much more complex. In general their values do depend on the current residual SNR and should be recomputed after each failed transmission. Consequently, it is not possible to simplify the problem in the same way as in CC and compute all targets at once for arbitrary $(l,L)$.
When $l=L-1$, the problem can be turned into a univariate, unconstrained optimization (by merit of $\epsilon_j^{(L)}=\frac{\epsilon_{tar}}{\prod_{i=0}^{L-1}\epsilon_j^{(i)}}$) and reads
\begin{mini}|s|[2]           	
    {\epsilon_j^{(L-1)}}           	
    {P_j^{(L-1)}\!+\!\int\displaylimits_{0}^{\gamma_{j}^{(L-1)}}\!f_e\left( x_L;\frac{P_j^{(L-1)}}{d_j^{\alpha}\sigma^2}\right)\!\left( -\frac{\frac{\gamma_{j}^{(L-1)} -x_L}{1+x_L} d_j^{\alpha}\sigma^2}{\ln(1-\epsilon_j^{(L)})} \right) dx_L}
    {\label{eq:optSimpleIR}}   	
    {\Psi_j^{(L-1,L)} (\gamma_j^{(L-1)},\Theta_j^{(L-1)}) =} 
\end{mini}
where the update to the residual SNR in incremental redundancy $\gamma_{ir_j}^{(l+1)} = \frac{\gamma_{ir_j}^{(l)} - \mathrm{SNR}_j^{(l)}}{1+\mathrm{SNR}_j^{(l)}}$ follows from the definition in \eqref{eq:errBIR}. The integral doesn't have a closed form, however a relatively simple approximation can be obtained by substituting exponential function with its first-order Taylor expansion around $0$ i.e. $e^{\frac{\ln(1-\epsilon_j^{(L-1)})x}{\gamma_j^{(L-1)}}}\approx \left( 1 + \frac{\ln(1-\epsilon_j^{(L-1)})}{\gamma_j^{(L-1)}}x \right)$. The approximated objective function becomes then
\begin{equation}
\begin{split}
-&\frac{\gamma_{j}^{(L-1)} d_j^{\alpha}\sigma^2}{\ln(1-\epsilon_j^{(L-1)})} + \frac{\ln(1-\epsilon_j^{(L-1)})d_j^{\alpha}\sigma^2}{\gamma_{j}^{(L-1)}\ln(1-\epsilon_j^{(L)})} \left( \frac{\gamma_{j}^{(L-1)}(\ln(1-\epsilon_j^{(L-1)})-2)}{2} \right. \\
& \left. + \ln(1-\epsilon_j^{(L-1)}) + \frac{(\gamma_{j}^{(L-1)}+1)(\gamma_{j}^{(L-1)}-\ln(1-\epsilon_j^{(L-1)}))\ln(\gamma_{j}^{(L-1)}+1)}{\gamma_{j}^{(L-1)}} \right)
\end{split}
\label{eq:approxIR}
\end{equation}

Since in this work we will consider only scenarios with at most $L=2$ retransmissions (in line with the low latency requirement) we adopt the following approach:
\begin{enumerate}
    \item For the few limited configurations characterized by transmission rate $R$ and $\epsilon_{tar}$ the optimal $\epsilon_j^{(0)}$, which is a solution to $\Psi_j^{(0,2)}~(2^R-~1,\epsilon_{tar})$ as defined in \eqref{eq:optGlob}, is found through an exhaustive search (performed offline). 
	To do this, we sweep through its possible values, fixing $\epsilon_j^{(0)}$, and then calculating the remaining expected power
	by solving and integrating \eqref{eq:approxIR} over a $[0,\gamma_j^{(0)}]$ range and with $\Theta_j^{(1)}=\frac{\epsilon_{tar}}{\epsilon_j^{(0)}}$.
	Note that since the initial $\gamma_j^{(0)}=2^R-1$ is identical for all UEs, so is the optimal $\epsilon_j^{(0)}$.
	\item After the first transmission, users who failed will end up with different residual SNRs. For each of them, the optimal error target for the upcoming retransmission is obtained separately by minimizing \eqref{eq:approxIR}.
	Since $\Psi_j^{(1,2)}~(\gamma_j^{(1)},\frac{\epsilon_{tar}}{\epsilon_j^{(0)}})$ is a univariate unconstrained problem with a closed form, it is relatively simple to obtain the solution numerically.
	\item In case second retransmission is necessary $\epsilon_j^{(2)}=\frac{\epsilon_{tar}}{\epsilon_j^{(0)}\epsilon_j^{(1)}}$ as follows from the constraint.
\end{enumerate}
\subsection{NOMA-HARQ}\label{sec:fading:noma}

As an enhancement of the OMA scheme we explore an approach in which the base station is allowed to schedule multiple UEs over the same channel resources. This can be useful especially in two instances 1) when due to the inherent randomness of new packet arrivals combined with decoding errors system enters a period of congestion and is forced to queue packets 2) when the residual SNR of unsuccessful packet is very low and assigning dedicated resources to a retransmission would be wasteful.\\
\indent Similarly to \cite{kotaba:icc} we start by defining the error probability of the transmission over shared channel resources. The expression is comprised of two terms: error probability with interferer's signal decoded and canceled, and the case when SIC is not successful. If the UE $j$, transmitting for the $l$-th time, is sharing the TF-block with UE $k$, who is currently at its $m$-th attempt, then 
\begin{equation}
\begin{split}
 p_{er_j}^{(l)} =& \Pr\left\lbrace \frac{P_j^{(l)} \left| h_j^{(l)} \right|^2}{d_j^{\alpha} \sigma^2} < \gamma_{j}^{(l)}, \frac{P_k^{(m)} \left| h_k^{(m)} \right|^2}{d_k^{\alpha} \left( \frac{P_j^{(l)} \left| h_j^{(l)} \right|^2}{ d_j^{\alpha}} \zeta_j + \sigma^2 \right)} > \gamma_{k}^{(m)}\right\rbrace \\
& + \Pr\left\lbrace \frac{P_j^{(l)} \left| h_j^{(l)} \right|^2}{d_j^{\alpha} \left( \frac{P_k^{(m)} \left| h_k^{(m)} \right|^2}{ d_k^{\alpha}}\zeta_k + \sigma^2 \right)} < \gamma_{j}^{(l)}, \frac{P_k^{(m)} \left| h_k^{(m)} \right|^2}{d_k^{\alpha} \left( \frac{P_j^{(l)} \left| h_j^{(l)} \right|^2}{ d_j^{\alpha}}\zeta_j + \sigma^2 \right)} < \gamma_{k}^{(m)}\right\rbrace
\end{split}
\label{eq:err2U}
\end{equation}
and the error probability of the second user of the TF-block $p_{er_k}^{(m)}$ is obtained by simply interchanging the indices $j\leftrightarrow k$ and $l \leftrightarrow m$. 
The coefficients $\zeta_j$ and $\zeta_k$ denote the interference reduction coefficients which will be explained later on.
This way of formulating \eqref{eq:err2U} is symmetrical and hence optimal in terms of ordering, i.e. if any of the two transmissions can be decoded in the presence of interference, the other one becomes interference-free. The expression \eqref{eq:err2U} can be obtained in the closed form as 
\begin{equation}
p_{er_j}^{(l)} = \begin{cases} \underbrace{1-  \left( \frac{S_j^{(l)}}{S_k^{(m)} \phi_k^{(m)}+S_j^{(l)}} 
+ \frac{S_k^{(m)}}{S_j^{(l)} \phi_j^{(l)} +S_k^{(m)}}e^{-\sigma^2\frac{\phi_j^{(l)}+1}{S_k^{(m)}}} \right) e^{-\frac{\sigma^2}{S_j^{(l)}}}}_{A}, & \mbox{if } \phi_j^{(l)}\phi_k^{(m)} \geq 1 \\ 
A - \left(1 - \frac{S_j^{(l)}}{S_k^{(m)}\phi_k^{(m)}+S_j^{(l)}} - \frac{S_k^{(m)}}{S_j^{(l)}\phi_j^{(l)}+S_k^{(m)}}  \right) e^{-\frac{\sigma^2}{1-\phi_j^{(l)}\phi_k^{(m)}}\left( \frac{\phi_j^{(l)}+1}{S_k^{(m)}} + \frac{\phi_k^{(m)}+1}{S_j^{l)}} \right)}, & \mbox{if } \phi_j^{(l)}\phi_k^{(m)} < 1 \\
\end{cases}
\label{eq:err2closed}
\end{equation}
where
$S_j^{(l)}=\frac{P_j^{(l)}}{\gamma_j^{(l)}d_j^{\alpha}}$ and $\phi_j^{(l)}=\gamma_j^{(l)}\zeta_j$. The derivation of \eqref{eq:err2closed} is discussed in Appendix \ref{sec:appendix:closed_form}.

Recall that in the OMA case, the first step was to find the optimal error target $\epsilon_j^{(l)}$ which then could be plugged into \eqref{eq:P1U} to determine the transmit power for the next transmission.
In the NOMA setting $\epsilon_j^{(l)}$ and $\epsilon_k^{(m)}$ that minimize the expected power per packet of each respective user would have to be found jointly which is significantly more difficult.
Due to its high computational complexity, in this work we will omit this process and instead use the same targets as for OMA. This can be further justified by analyzing the results and findings presented in \cite{kotaba:icc} which show that the optimal error targets for OMA and NOMA are in fact very similar.

With $\epsilon_j^{(l)},\epsilon_k^{(m)}$ given and the error probabilities defined as in \eqref{eq:err2closed}, the transmit powers are assigned to users by solving the following optimization problem:
\begin{argmini!}|s|[2]           	
    {P_j^{(l)},P_k^{(m)}}           	
    {P_j^{(l)} + P_k^{(m)} \label{eq:NOMARaylObj1}}   
    {\label{eq:optNOMARayl}}   	
    {}                                
    \addConstraint{p_{er_j}^{(l)}\leq \epsilon_j^{(l)},\quad}{p_{er_k}^{(m)} \leq \epsilon_k^{(m)} \label{eq:NOMARaylCon1}}    
\end{argmini!}
The solution is found using an interior-point convex solver.
While the constraint functions are not convex, the domain can be divided into two disjoint regions: $\frac{P_j^{(l)}}{d_j^{\alpha}} > \frac{P_k^{(m)}}{d_k^{\alpha}}$ and $\frac{P_j^{(l)}}{d_j^{\alpha}} < \frac{P_k^{(m)}}{d_k^{\alpha}}$. 
The global minimum is determined by finding the local minimum of each and selecting the lower one.\looseness=-1

\subsubsection{Chase Combining}\label{sec:fading:noma:cc} 
When using NOMA with Chase Combining additional assumption is required for the expression \eqref{eq:err2U} to be valid. Note that the total SINR of a packet can be written as a simple sum of the SINRs of its individual copies only when the interference in each of them is uncorrelated. For that reason we ensure in our simulations that throughout its $L+1$ transmissions a packet is never paired more than once with the same packet of other user.
This is rarely an issue and does not impact reliability, only the scheduling process explained later on.

In certain cases the procedure described in \cite{kotaba:icc} can be refined by utilizing the previous copies of the interfering packet to partially cancel its contribution in the current transmission even before applying the SIC. This is reflected in \eqref{eq:err2U} by the reduction coefficients $\zeta_j$ and $\zeta_k$. 
The details on how to obtain them can be found in Appendix \ref{sec:appendix:reduction}.

\subsubsection{Incremental Redundancy}\label{sec:fading:noma:ir} 
In addition to CC-HARQ, we investigate the NOMA approach with IR. Since in IR each packet is composed of different symbols, it can be assumed that all of them  experience independent interference. As a result, there is no need for the additional constraint on the scheduling that was required for CC. At the same time, since the additional interference reduction in CC was achieved by combining previous signals containing the same packet it is no longer possible to use this feature with IR
\footnote{An equivalent technique could be attempted with incremental redundancy, however the exact procedure and resulting gains are difficult to asses. To suppress the current interfering packet the previous (unsuccessful) packets would have to be soft-decoded and then re-encoded with mother code rate to ``guess'' the next corresponding symbols in the buffer.} 
and $\zeta_j,\zeta_k=1$.

\section{Instantaneous CSI and finite blocklength}\label{sec:finite}
The preceding analysis pertained to the case of Rayleigh fading channel whose realizations are unknown until after the reception of the packet (through perfect estimation) and a priori only their distribution is known. 
However, due to the low end-to-end latency of the URLLC communication it is of interest to investigate also the case where the channel coherence time is large enough that the BS can treat the channel during subsequent uplink transmission as known.
Unlike in statistical CSI case where the dominant source of errors is fading \cite{finite:vs:asymptotic}, here the finite blocklength effects become crucial. Since the channel effectively becomes AWGN, the decoding errors are caused solely by noise which is especially prominent in short packets. Hence, to study the case of instantaneous CSI we resort to finite blocklength analysis \cite{finite} and, for tractability reasons, limit the scope to just the case of IR-HARQ.

\subsection{Finite blocklength OMA-HARQ}\label{sec:finite:oma}
As previously, let us start with a simpler case of dedicated resources. The average mutual information contained in $l$-th transmission of the packet from user $j$ can be written as \cite{finite}
\begin{equation}
\begin{aligned}
I_j^{(l)} = & \frac{1}{K}\sum_{n=1}^K \left[ \ln \left(1+\frac{P_j^{(l)} \left| h_j^{(l)} \right|^2}{d_j^{\alpha} \sigma^2} \right) + \frac{\left| y_{j,n}^{(l)} \right|^2}{\frac{P_j^{(l)} \left| h_j^{(l)} \right|^2}{d_j^{\alpha}} + \sigma^2} - 
 \frac{\left| y_{j,n}^{(l)} - \sqrt{\frac{P_j^{(l)}}{d_j^{\alpha}}}h_j^{(l)}x_{j,n}^{(l)} \right|^2}{\sigma^2} \right]
\end{aligned}
\label{eq:mut}
\end{equation}
where the sum involves all received and transmitted symbols $y_{j,n}^{(l)}$ and $x_{j,n}^{(l)}$ respectively. In \eqref{eq:mut} the difference of the last two terms is a Laplacian random variable with zero mean and variance equal to $\frac{2P_j^{(l)} \left| h_j^{(l)} \right|^2}{P_j^{(l)} \left| h_j^{(l)} \right|^2 + d_j^{\alpha}\sigma^2}$. As shown in \cite{finite} a sum of $K$ such Laplacian random variables can be well approximated by a Gaussian random variable with zero mean and $K$ times higher variance. Hence, the average mutual information contained in a codeword of size $K$ follows:
\begin{equation}
\begin{aligned}
\hat{I}_j^{(l)} \sim \mathcal{N} & \left( \ln \left(1+\frac{P_j^{(l)} \left| h_j^{(l)} \right|^2}{d_j^{\alpha} \sigma^2} \right), \frac{2P_j^{(l)} \left| h_j^{(l)} \right|^2}{K\left(P_j^{(l)} \left| h_j^{(l)} \right|^2 + d_j^{\alpha} \sigma^2\right)} \right)
\end{aligned}.
\label{eq:Iapprox}
\end{equation}
Since the codewords in IR can be treated as independent, the total mutual information provided by $l$ subsequent transmissions is also a Gaussian random variable with mean $\mu_j^{(l)} =  \sum_{i=0}^{l} \ln \left(1+\frac{P_j^{(i)} \left| h_j^{(i)} \right|^2}{d_j^{\alpha} \sigma^2} \right)$ and variance $\nu_j^{(l)} = \frac{1}{K}  \sum_{i=0}^{l} \frac{2P_j^{(i)} \left| h_j^{(i)} \right|^2}{P_j^{(i)} \left| h_j^{(i)} \right|^2 + d_j^{\alpha} \sigma^2}$ .

Again, the ultimate goal is to minimize the expected total power per packet, however the optimization problem is considerably different. 
Since the immediate channel realization is known, it is clear that the optimal transmit power (and the corresponding optimal error probability)
is a function of both the instantaneous channel gain and the statistics of the future channel realizations. 
Moreover, after the failed attempt, the receiver is not able to determine the exact residual mutual information\footnote{In fact, the residual mutual information $R\ln{2} - \sum_{i=0}^{l} \hat{I}_j^{(i)}$ is a random variable following truncated Gaussian distribution restricted to $[0,R\ln{2}]$.} as it depends on the particular realizations of the noise which are unknown. 
To determine the transmit power for the packet at round $l$ belonging to user $j$, the BS needs to solve the recursive optimization problem which can be framed as follows:
\begin{mini!}|s|[2]           	
    {P_j^{(l)}}           	
    {P_j^{(l)} + \epsilon_j^{(l)} \int_{0}^{\infty} e^{-z_{l+1}} \Psi_{j}^{(l+1,L)} \left( z_{l+1},\mu_j^{(l)},\nu_j^{(l)} \right) d z_{l+1} \label{eq:globObjShort1}}   
    {\label{eq:optGlobShort}}   	
    {\Psi_{j}^{(l,L)} \left( \left|h_j^{(l)} \right|^2,\mu_j^{(l-1)},\nu_j^{(l-1)} \right) =  }                                
    \addConstraint{F_{\mathcal{N}}\left(R \ln 2; \mu_j^{(L)}, \nu_j^{(L)} \right)}{\leq \epsilon_{tar} \label{eq:globConShort1}}  
\end{mini!}
where $F_{\mathcal{N}}(\cdot;\cdot,\cdot)$ is the CDF of Gaussian distribution, $\epsilon_j^{(l)} = \frac{ F_{\mathcal{N}}\left(R \ln 2; \mu_j^{(l)}, \nu_j^{(l)} \right)}{F_{\mathcal{N}}\left(R \ln 2; \mu_j^{(l-1)}, \nu_j^{(l-1)} \right)}$ is the failure probability and the integration is over the possible channel gains in the next uplink phase. 
Note that both $\epsilon_j^{(l)}$ and the recursive term $\Psi_{j}^{(l+1,L)} \left( z_{l+1},\mu_j^{(l)},\nu_j^{(l)} \right)$ depend on the latest $P_j^{(l)}$ and $\left|h_j^{(l)} \right|^2$  as they are included in $\mu_j^{(l)}$ and $\nu_j^{(l)}$.

As was the case earlier, the general closed form expression for the objective function \eqref{eq:globObjShort1} is difficult to obtain. In the last attempt, i.e. $l = L$, the optimal power follows directly from the constraint \eqref{eq:globConShort1}. More specifically, $P_j^{(L)} = \rho \frac{d_j^{\alpha}}{\left| h_j^{(L)} \right|^2}$, where $\rho$ is the solution to the equation
\begin{equation}
\frac{1}{2}\left(1+ \mathrm{erf} \left( \frac{R\ln 2 - \ln \left(1+\frac{\rho}{\sigma^2} \right) - \mu_j^{(L-1)}}{ \sqrt{\frac{4\rho}{K\left( \rho + \sigma^2 \right)} + \nu_j^{(L-1)}}}\right) \right) = \epsilon_{tar}
\label{eq:shortUlt}
\end{equation}
with $\mathrm{erf}(\cdot)$ being the error function. For $l = L-1$, the optimal power $P_j^{(L-1)}$ can be determined by minimizing the objective \eqref{eq:globObjShort1} which in that case becomes
\begin{equation}
P_j^{(L-1)} + \frac{ F_{\mathcal{N}}\left(R \ln 2; \mu_j^{(L-1)}, \nu_j^{(L-1)} \right)}{F_{\mathcal{N}}\left(R \ln 2; \mu_j^{(L-2)}, \nu_j^{(L-2)} \right)}\ \rho \int_{0}^{\infty} \frac{1}{z_L} e^{-z_L} d z_L  .
\label{eq:shortPenUlt}
\end{equation}
The integral, which represents the expected value of $1/\left| h_j^{(L)} \right|^2$ (inverse exponential distribution), does not converge. To remedy that, we assume that any packet which is in a deep enough fade during its last $L$-th transmission will be dropped. Since the overall target error rate is $\epsilon_{tar}$ we select $\epsilon_{drop} < \epsilon_{tar}$ and find the point through inverse CDF $\left|h_{fade}\right|^2 = F_e^{-1}\left(\epsilon_{drop};1 \right)$ which will be the lower limit for the integration. 
Although minimization of \eqref{eq:shortPenUlt} requires solving recursively \eqref{eq:shortUlt} as well (choice of $P_j^{(L-1)}$ determines $\rho$), it can still be done quite efficiently numerically.

\begin{figure*}[t!]%
\centering
\subfigure[]{
\label{fig:trial1}%
\includegraphics[width=0.33\textwidth]{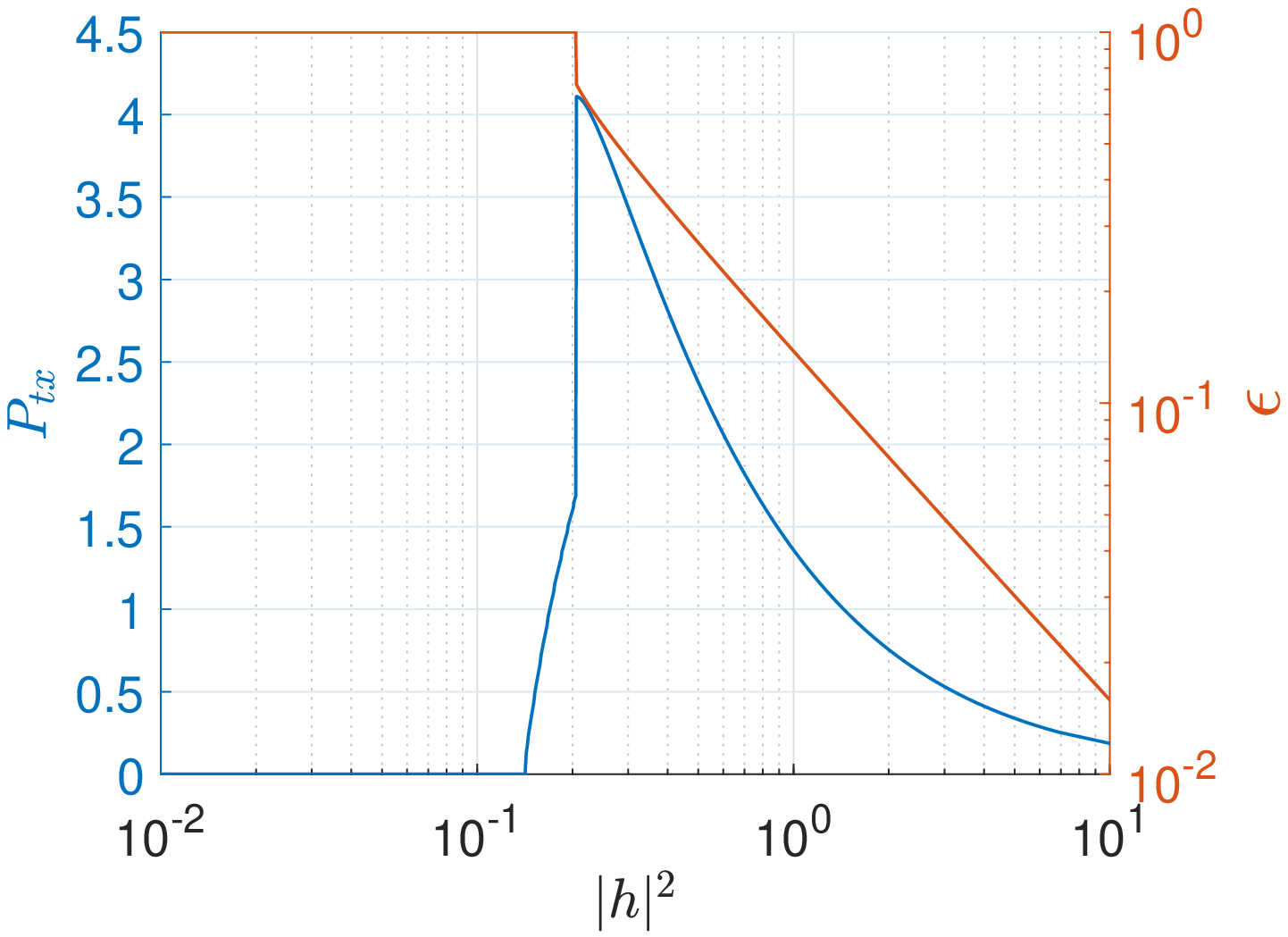}}%
\hfill
\subfigure[]{
\label{fig:trial2}%
\includegraphics[width=0.33\textwidth]{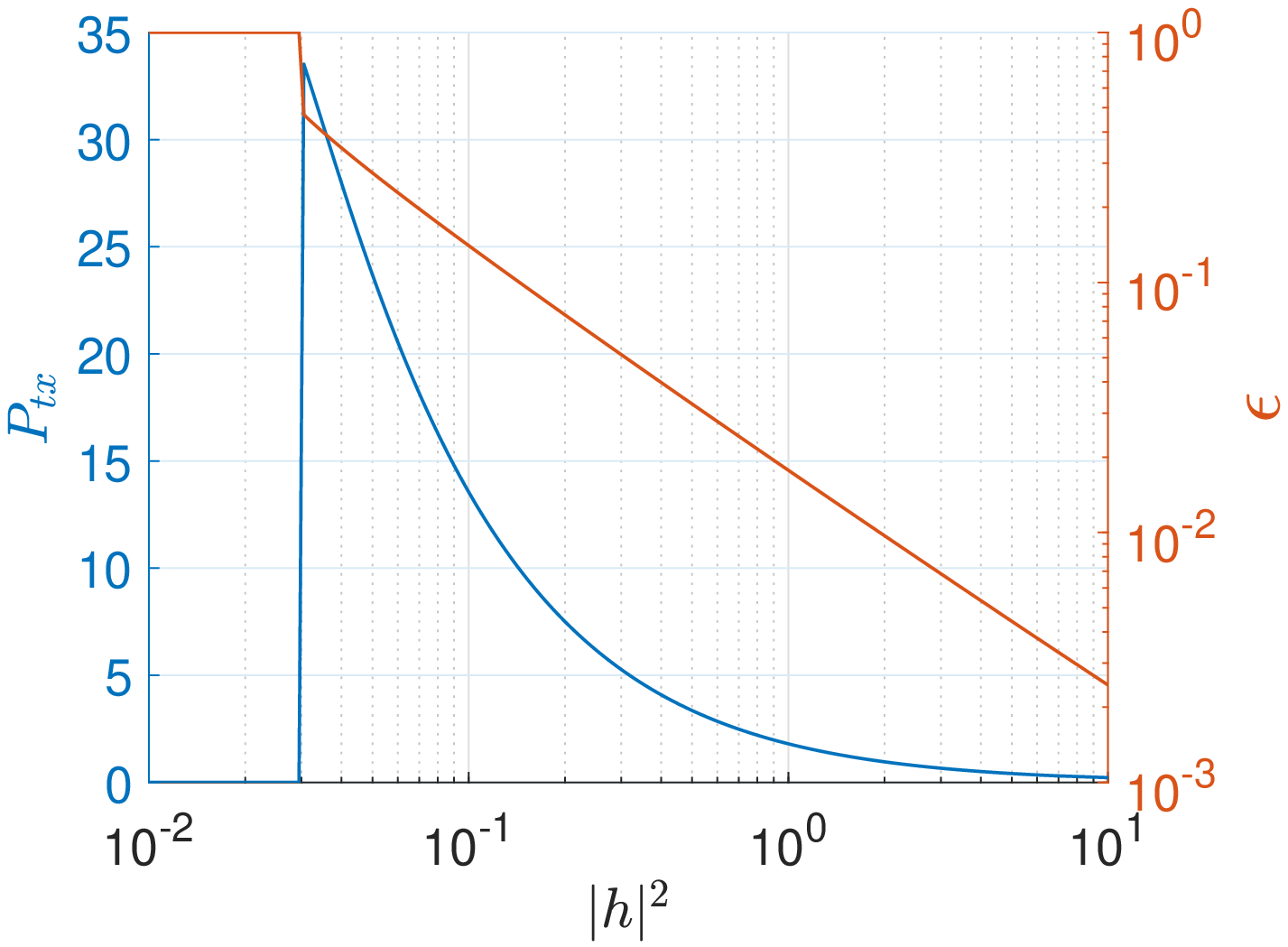}}%
\hfill
\subfigure[]{
\label{fig:trial3}%
\includegraphics[width=0.33\textwidth]{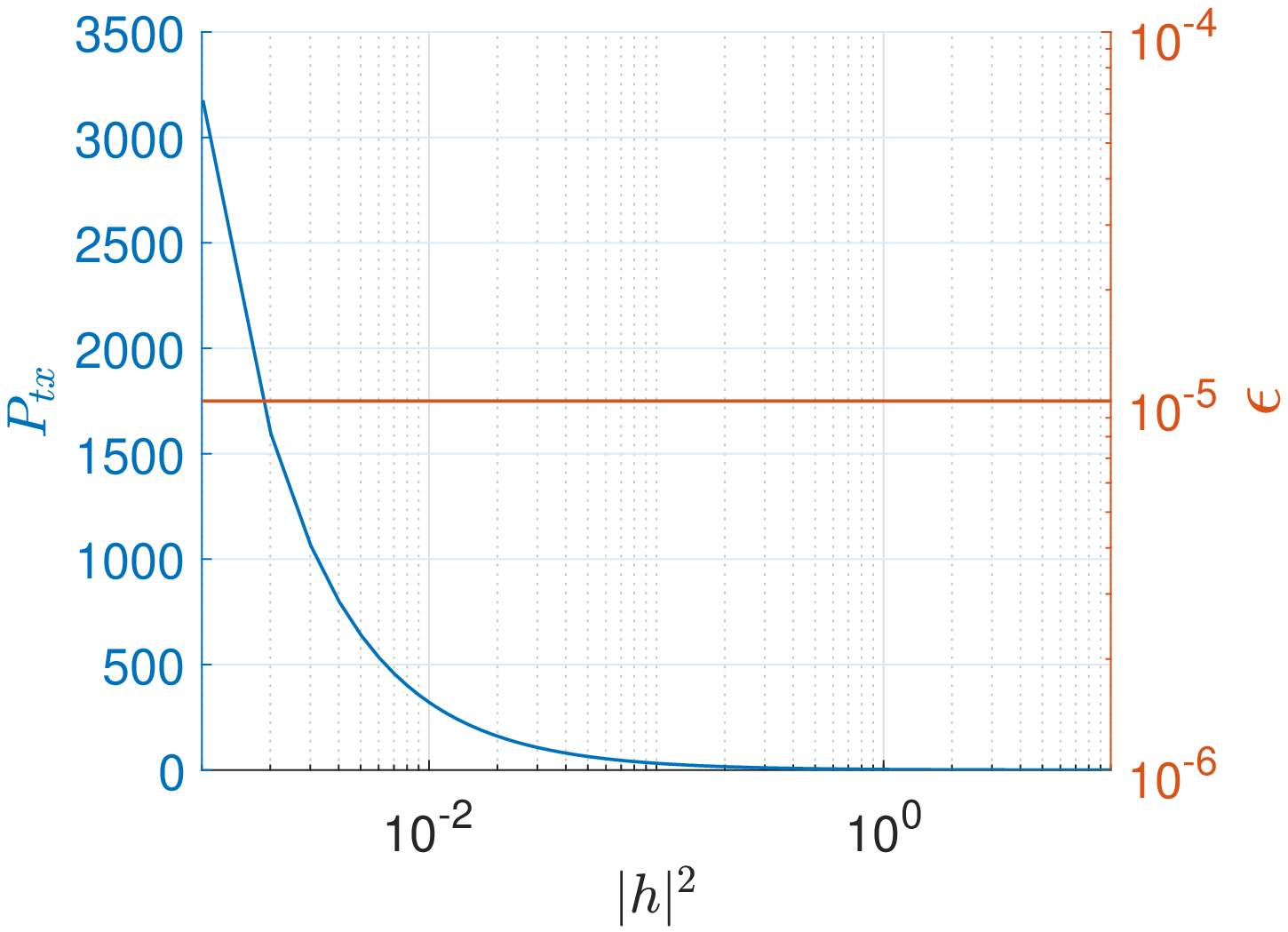}}%
\caption{The optimal transmit power and resulting error probability as a function of current channel gain realization for $R=1$. The values are normalized to $\sigma^2=1$ and $d^{\alpha} = 1$. The three figures correspond to the power used during initial transmission assuming (a) $L=2$, (b) $L=1$, (c) $L=0$ allowed retransmissions.}
\label{fig:powerVSchannel}
\end{figure*}

For $l \leq L-2$ the optimization becomes even more complex as it adds another level of recursion and since in this work we consider at most $L=2$ retransmissions, $\Psi_j^{(0,2)}$ can become a bottleneck. A better idea is to precompute $P_j^{(0)}$ as a function of $\left|h_j^{(0)}\right|^2$ offline which we show in Fig.\ref{fig:powerVSchannel}. In practice, during our simulations the following approach is used:
\begin{enumerate}
	\item For the newly arrived packet experiencing $\left| h_j^{(0)} \right|^2$, the optimal power $P_j^{(0)}$ is determined by interpolation on Fig.~\ref{fig:trial1}.
	\item If the packet fails during initial transmission, the optimal power $P_j^{(1)}$ is determined by minimizing \eqref{eq:shortPenUlt}. If the packet ended up in round 1 due to being postponed (e.g. because of the deep fade) then $P_j^{(1)}$ can be interpolated from Fig.~\ref{fig:trial2}. Note that this is a consequence of $\Psi_j^{(0,1)}$ being equivalent to $\Psi_j^{(1,2)}$ with $P_j^{(0)}=0$.
	\item If the packet fails for the second time, then $P_j^{(2)}$ is obtained by solving \eqref{eq:shortUlt}.
\end{enumerate}

Lastly, we note that the optimal power obtained by solving \eqref{eq:optGlobShort} can be $0$ (cf. Fig. \ref{fig:powerVSchannel}). This means the BS consciously chooses to postpone the packet based on unfavourable $\left|h_j^{(l)}\right|^2$.

\subsection{Finite blocklength NOMA-HARQ}\label{sec:finite:noma}

Lastly, we move on to discuss the application of NOMA-HARQ in the finite blocklength scenario. Let $P_j^{(l)}$ and $P_k^{(m)}$ be the transmit powers of the two packets which are to be scheduled in the same TF-block and let the packet from user $j$ be attempted to decode first\footnote{Accounting for optimal SIC ordering becomes relevant when there is an uncertainty about the relationship between two received powers. Clearly this is not the case when channel realizations are known in advance.}. We will also denote the received powers as $Q_j^{(l)}=\frac{P_j^{(l)} \left| h_j^{(l)} \right|^2}{d_j^{\alpha}}$, $Q_k^{(m)}=\frac{P_k^{(m)} \left| h_k^{(m)} \right|^2}{d_k^{\alpha}}$ respectively. The resulting error probabilities are
\begin{equation}
p_{er_j}^{(l)} = \frac{ F_{\mathcal{N}}\left(R \ln 2; \ln \left(1+\frac{Q_j^{(l)}}{Q_k^{(m)} + \sigma^2 } \right) + \mu_j^{(l-1)}, \frac{2Q_j^{(l)}}{K \left( Q_j^{(l)} + Q_k^{(m)} + \sigma^2 \right)} + \nu_j^{(l-1)} \right)}{F_{\mathcal{N}}\left(R \ln 2; \mu_j^{(l-1)}, \nu_j^{(l-1)} \right)}
\end{equation}

\begin{equation}
\begin{aligned}
p_{er_k}^{(m)} &= \left( 1 - p_{er_j}^{(l)} \right) \frac{ F_{\mathcal{N}}\left(R \ln 2; \ln \left(1+\frac{Q_k^{(m)}}{\sigma^2} \right) + \mu_k^{(m-1)}, \frac{2Q_k^{(m)}}{K \left( Q_k^{(m)} + \sigma^2 \right)} + \nu_k^{(m-1)} \right)}{F_{\mathcal{N}}\left(R \ln 2; \mu_k^{(m-1)}, \nu_k^{(m-1)} \right)}\\
& + p_{er_j}^{(l)} \frac{ F_{\mathcal{N}}\left(R \ln 2; \ln \left(1+\frac{Q_k^{(m)}}{ Q_j^{(l)} + \sigma^2 } \right) + \mu_k^{(m-1)}, \frac{2Q_k^{(m)}}{K \left( Q_k^{(m)} + Q_j^{(l)} + \sigma^2 \right)} + \nu_k^{(m-1)} \right)}{F_{\mathcal{N}}\left(R \ln 2; \mu_k^{(m-1)}, \nu_k^{(m-1)} \right)}\\
\end{aligned}
\end{equation}

We make note here of the slight abuse of our usage of the terms $\mu$ and $\nu$. These are meant to represent the means and variances of the mutual information obtained in earlier rounds and are clearly a function of the SINR of the signal of interest. Since in the NOMA approach earlier replicas of the packet could have also been scheduled non-orthogonally one should keep in mind to include the appropriate interference terms in the calculations of $\mu$ and $\nu$.

To find the appropriate powers for the two UEs we follow similar heuristic as before. 
Let ${P_j^{(l)}}^{\star}$ and ${P_k^{(m)}}^{\star}$ be the optimal OMA powers of users $j$ and $k$ for the upcoming round as given by \eqref{eq:optGlobShort}. Consequently, in the OMA setting, these powers would result in the error probabilities ${\epsilon_j^{(l)}}^{\star}$ and ${\epsilon_k^{(m)}}^{\star}$ respectively. Then, the goal is to find appropriate $P_j^{(l)}$ and $P_k^{(m)}$ for the NOMA transmissions such that the individual OMA error targets are met:
\begin{argmini!}|s|[2]           	
    {P_j^{(l)},P_k^{(m)}}           	
    {P_j^{(l)} + P_k^{(m)} \label{eq:NOMAshortObj1}}   
    {\label{eq:optNOMAshort}}   	
    {}                                
    \addConstraint{p_{er_j}^{(l)} \leq {\epsilon_j^{(l)}}^{\star},\quad}{p_{er_k}^{(m)} \leq {\epsilon_k^{(m)}}^{\star} \label{eq:NOMAshortCon1}}    
\end{argmini!}

The rationale behind using the same error targets for NOMA as for OMA is again the tractability of the problem.
Finding even a single pair of optimal NOMA targets has high computational complexity, and in the instantaneous CSI case they would have to be computed for each pair of values $\Big(\left| h_j^{(l)} \right|^2, \left| h_k^{(m)} \right|^2\Big)$.
The solution to \eqref{eq:optNOMAshort} is found numerically.

\section{Scheduling}\label{sec:scheduling}

The last element missing before we move on to the results is the matter of scheduling. As already mentioned, in this work we consider a system having a finite amount of resources, namely $W$ TF-blocks. These might not be enough to accommodate all the packets of the active users which inevitably leads to queuing and requires defining a scheduling policy. Because of the complexity of the problem whose optimal solution would require taking into account both current packets and future arrivals and since scheduling is not the primary topic of this work, we decide to settle for a heuristic approach which will be now described.

\subsection{OMA scheduling}\label{sec:scheduling:oma}

When the total number of packets in the system $T$ is lower than the amount of resources $W$ the scheduling decision is straightforward. Furthermore, in the instantaneous CSI case, the BS can already at this point decide that some of the packets should be postponed based on their poor channel conditions. 
We also remark that there is no limit regarding how many of them a single UE can send in one UL phase as long as there are available resources and is instructed to do so by the BS\footnote{Because at each step user generates a new packet with probability $b$ and, since the allowed number of retransmissions is $L$, each user can be storing in its buffer up to $L+1$ packets at any given time (each at a different round).}.

In case the number of packets $T$ exceeds $W$, the BS performs an intermediate step and decides which of them should be postponed. To do this we adopt the following procedure:
\begin{enumerate}
\item The priority is given to the packets currently in their last $L$-th round. If the number $T_C$ of such critical packets exceeds $W$ the ones which require the least power are transmitted and the remaining are dropped. All the non-critical $T-T_C$ packets are postponed, i.e. they are moved to the next round. Note that dropping the packets is the last resort since it compromises the overall reliability.
\item If $T_C < W$, the remaining TF-blocks are used to transmit some of the non-critical packets.
For each of them BS calculates two values:
current expected OMA power $\Psi_j^{(l,L)}$, and the expected power assuming this round was skipped $\Psi_j^{(l+1,L)} |P_j^{(l)}=0$. 
Note that since $P_j^{(l)}=0 \implies \epsilon_j^{(l)}=1$ the second value entails more aggressive future error targets that will account for the lower number of retransmission opportunities.
The packets that will be scheduled are those with the highest difference $(\Psi_j^{(l+1,L)} |P_j^{(l)}=0) - \Psi_j^{(l,L)}$. 
The rationale is to prioritize the packets which are the most ``expensive'' to postpone.
Depending on the CSI scenario, the expected powers are obtained based on either \eqref{eq:optGlob} or \eqref{eq:optGlobShort}.
\end{enumerate}
Once the set of packets that will be sent is established,
their optimal transmit powers are determined as outlined in the appropriate section (statistical/instantaneous CSI, CC/IR).

\subsection{NOMA scheduling}\label{sec:scheduling:noma}
The overall procedure for deciding which packets to postpone is similar to the OMA case with the following caveats:
\begin{enumerate}
\item Since pairing allows to accommodate twice as many packets, queuing starts only when $T>2W$. Again, the priority is given to the packets at round $L$ and if $T_C > 2W$ the ones requiring highest power are dropped.
\item When deciding which of the non-critical packets to transmit immediately and which to postpone the procedure is identical as for OMA, i.e. the calculation of the expected powers is also based on the equations derived for OMA. While this is a suboptimal approach, it is not clear how to compare the current and future NOMA powers since that would require the a priori knowledge of which users will be active next and what will be the exact pairing now and in the future. Instead, we resort to a simple heuristic that if the packet is ``expensive'' to postpone in OMA terms, then it is also the case for NOMA, especially since the latter always require some extra power.
\end{enumerate}

The next step is to determine the pairing.
Let us denote the number of pairs to be created as $q$. When $T \geq 2W$ then $q = W$, however when $T<2W$ we can consider two cases for our simulations. a) Power conservative (PC) approach which will form pairs only if necessary, i.e. when $T>W$, leading to $q=\max (T-W,0)$. Consequently the usage of resources is maximized.
b) Resource conservative (RC) approach where as many pairs as possible are made resulting in $q = \left \lfloor \frac{T}{2} \right \rfloor $. The usage of resources is minimized at the cost of higher power.

Once it is decided which packets will be transmitted and how many pairs are needed, the appropriate matching of the users is determined:
\begin{enumerate}
\item First, for each pair of packets we calculate the optimal NOMA powers according to either \eqref{eq:optNOMARayl} or \eqref{eq:optNOMAshort}. Then, we compare them with the optimal OMA powers of each of the user to determine the difference $\left(P_{j,NOMA}^{(l)} + P_{k,NOMA}^{(m)} \right) - P_{j,OMA}^{(l)} - P_{k,OMA}^{(m)}$, which is the extra cost of scheduling the two packets together rather than on dedicated resources. Note that some of the pairs cannot be formed. The possible reasons include: earlier joint transmission (only applicable to CC-HARQ), optimization \eqref{eq:optNOMARayl} or \eqref{eq:optNOMAshort} did not converge or the two packets belong to the same UE.
\item Once the costs for all pairs are known, $q$ pairs that produce the lowest combined cost are selected. This step is a variation of the maximum weight matching problem which can be solved by Blossom algorithm \cite{blossom1}. 
\end{enumerate}

\section{Results}\label{sec:res}
The parameters used for simulations are gathered in Table \ref{tab1}.
The power of noise is given per symbol and is calculated as $\sigma^2=N_0+10\log_{10}B_w$, where $N_0=-173.9\mathrm{dBm/Hz}$ is a typical power spectral density at $298K$ and $B_w=30\mathrm{kHz}$ was chosen as the symbol bandwidth. 
Throughout this section and in the legends of the figures we will refer to the Chase combining and incremental redundancy HARQ utilizing statistical CSI knowledge as CC and IR respectively, while the IR-HARQ with instantaneous CSI and finite blocklength as Finite IR.
Furthermore, we will distinguish three access methods: OMA, power conservative NOMA (PC-NOMA) and resource conservative NOMA (RC-NOMA).
\begin{table}[t!]
 \renewcommand{\arraystretch}{1.4}
 \centering 
 \caption{Simulation parameters}
 \begin{tabular}{| c | c | c | c |}
    \hline
    Number of UEs $N$ & $40$ & Number of symbols $K$ & $50$\\ \hline
    Number of TF-blocks $W$ & $10$ & Final BLER $\epsilon_{tar}$ & $10^{-5}$\\ \hline
    Number of retransmissions $L$ & $2$ & Deep fade threshold $\epsilon_{drop}$ & $10^{-6}$\\ \hline
    Min. distance $D_{min}$ & $20$ m & Activation probability $b$ & $b \in [0.05,0.5]$\\ \hline
    Max. distance $D_{max}$ & $120$ m & Transmission rate $R$ & $R\in [0.5, 2.5]$ bits/symbol\\ \hline
    Pathloss exponent $\alpha$ & $2$ & Channel Type & Rayleigh block fading\\ \hline
    Noise power $\sigma^2$ & $-129.1$ dBm & Channel estimation method & Perfect\\ \hline
  \end{tabular}
  \label{tab1}
\end{table}
\begin{table}[t!]
 \renewcommand{\arraystretch}{1.4}
 \centering 
 \caption{Optimal error targets in statistical CSI HARQ and $\epsilon_{tar}=10^{-5}$}
 \begin{tabular}{| c | c | c | c | c | c | c |}
    \hline
     CC & IR, $R=0.5$ & IR, $R=1$ & IR, $R=1.5$ & IR, $R=2$ & IR, $R=2.5$\\ \hline
    $\epsilon^{(0)}=0.189$, $\epsilon^{(1)}=0.0374$, $\epsilon^{(2)}=0.0014$ & $\epsilon^{(0)}=0.2$ & $\epsilon^{(0)}=0.215$ & $\epsilon^{(0)}=0.2255$ & $\epsilon^{(0)}=0.2485$ & $\epsilon^{(0)}=0.262$\\ \hline
  \end{tabular}
  \label{tab:errs}
\end{table}

In Table \ref{tab:errs} we provide the optimal error targets calculated for IR and CC.
In general, the initial error target $\epsilon^{(0)}$ in IR is more relaxed and, unlike in CC, it increases with the transmission rate. Consequently, when using IR, the higher the rate, the more the system will rely on retransmissions to achieve the required reliability.
\begin{figure}[t!]%
\centering
\subfigure[OMA]{
\label{fig:avail:oma}%
\includegraphics[width=0.42\textwidth]{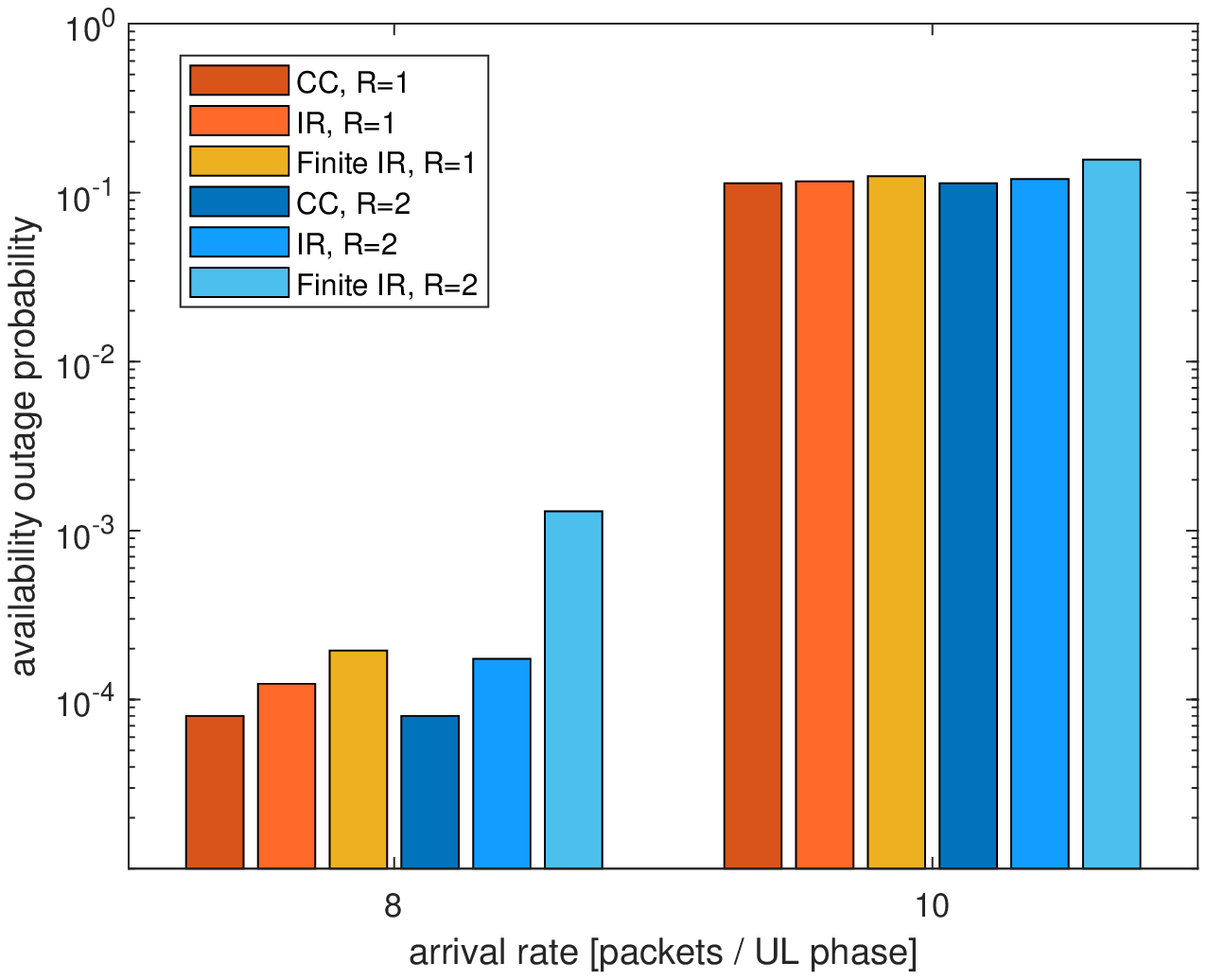}}%
\hfill
\subfigure[NOMA]{
\label{fig:avail:noma}%
\includegraphics[width=0.42\textwidth]{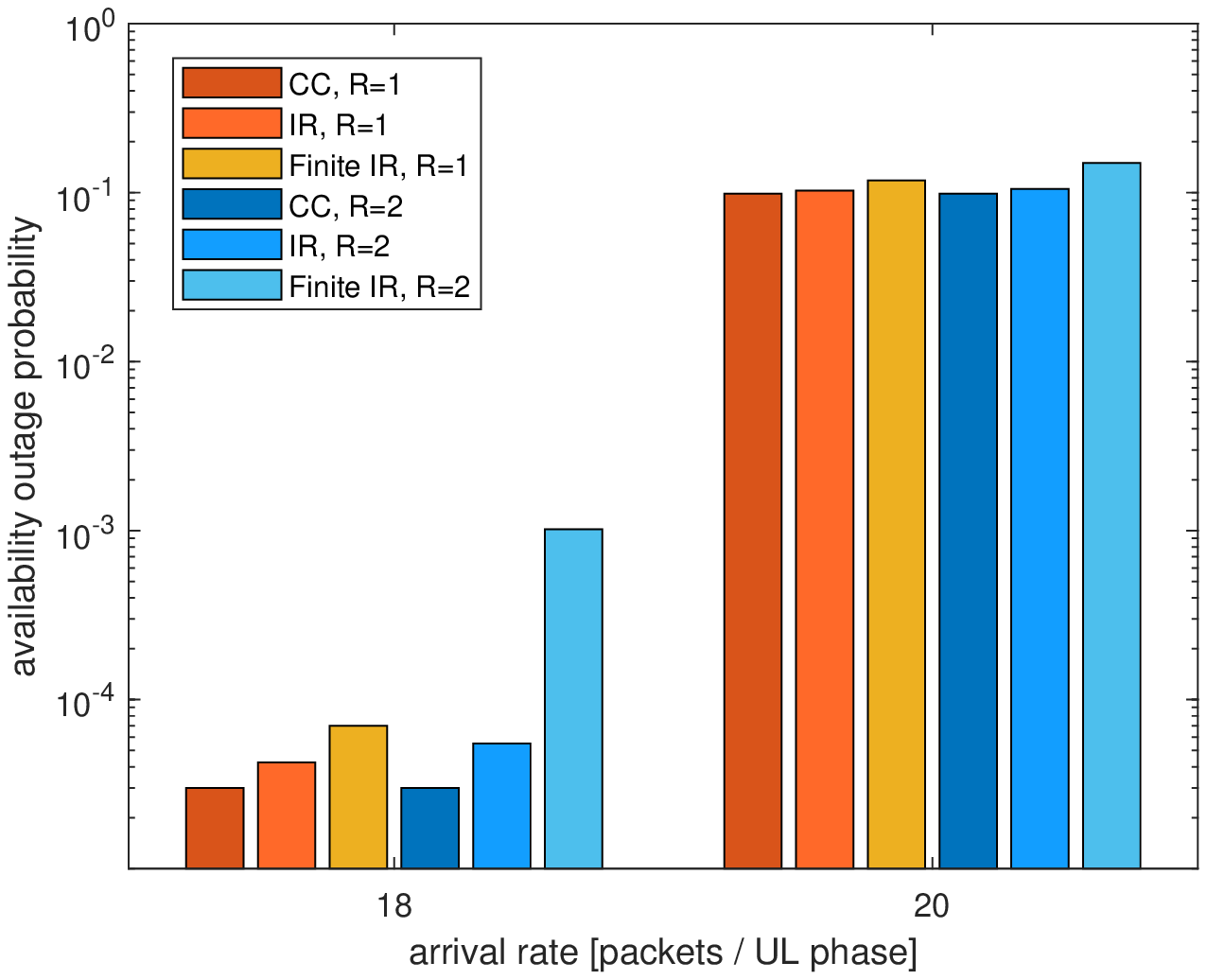}}%
\caption{The availability outage probability for (a) OMA and (b) NOMA.}
\label{fig:res:1}
\end{figure}

Let us start by looking into the most fundamental difference between OMA and NOMA which is reflected in their availability outage performance. We define availability outage as the state in which BS is forced to drop packets (i.e. timeslots where $T_C > 2W$). This way we make a clear distinction between availability and reliability similarly to \cite{3gpp:availability}. Note that all packets which are not dropped have their reliability requirements fulfilled, as this is ensured by the power optimization and selection step.
Fig.~\ref{fig:res:1} depicts the availability of OMA and NOMA system as a function of the mean number of new packets per uplink phase $bN$.
For arrival rates which are below the shown values ($bN<8$ for OMA and $bN<18$ for NOMA) the availability outage probability becomes much lower than the transmission outage probability of $10^{-5}$.
Conversely, arrival rates higher than $W$ and $2W$ result in an unstable system.
In terms of availability PC-NOMA and its RC variant perform almost identically, hence, for brevity, only the former is presented.
This is due to the fact that availability becomes an issue only as the mean number of arrivals approaches the system bandwidth, at which point PC and RC methods become equivalent since $T\geq2W$ most of the time\footnote{Note that the total number of packets $T$ is a sum of new arrivals, postponed packets and those that failed previous transmission.}.
In this example the introduction of NOMA allows to support URLLC traffic of more than two times higher intensity compared to the baseline OMA.
For a given arrival rate, the differences in availability outage between the three methods are a consequence of their distinct error targets for the initial transmission $\epsilon_j^{(0)}$, which are most demanding for CC, and least for Finite IR. Furthermore, they also increase with rate $R$ (except for CC).
Since retransmissions add up to an already high number of new packets, when using CC the probability of driving the system into availability outage is lowest.

\begin{figure}[t!]%
\centering
\subfigure[]{
\label{fig:ppp:cc}%
\includegraphics[width=0.33\textwidth]{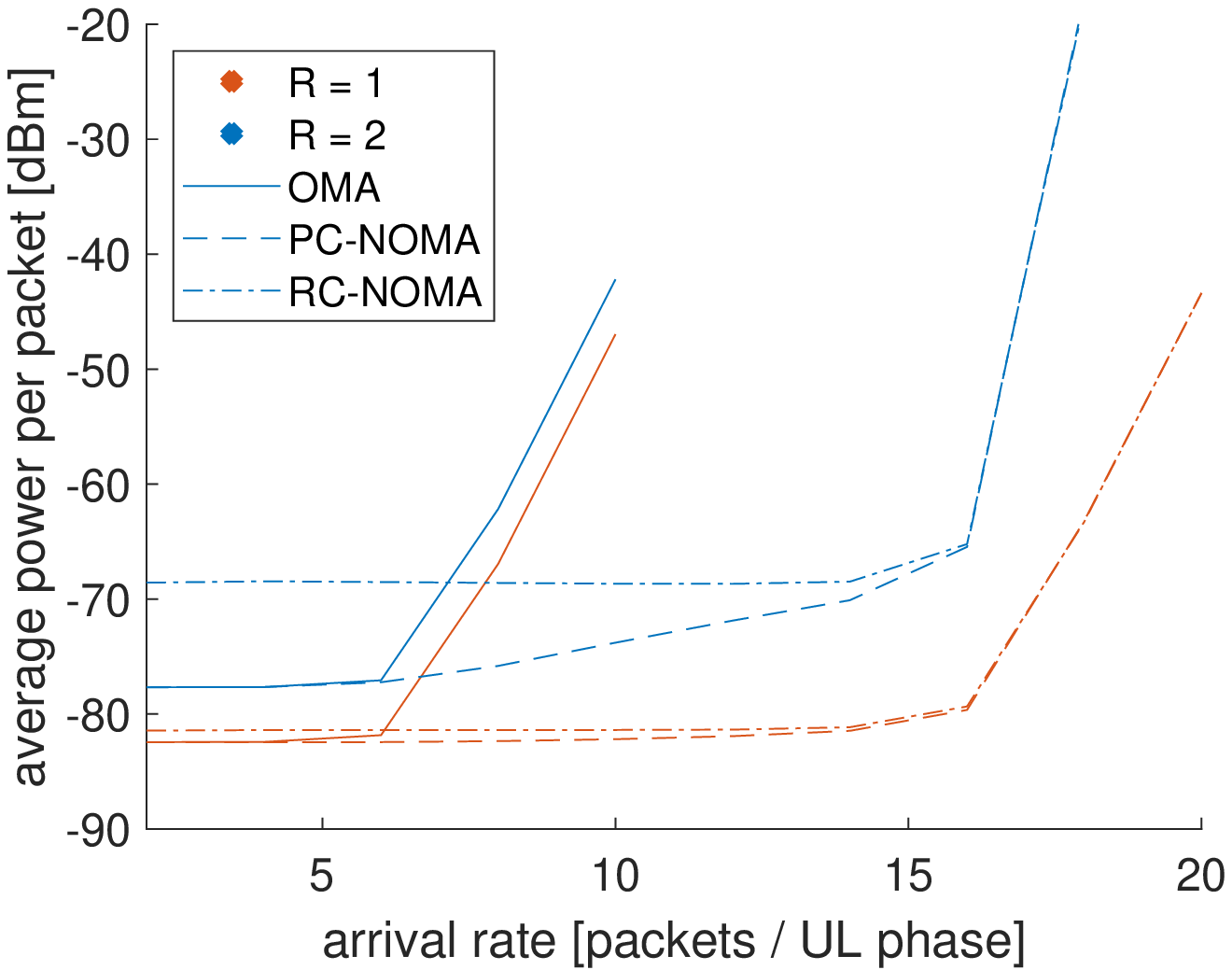}}%
\hfill
\subfigure[]{
\label{fig:ppp:cc_vs_ir}%
\includegraphics[width=0.33\textwidth]{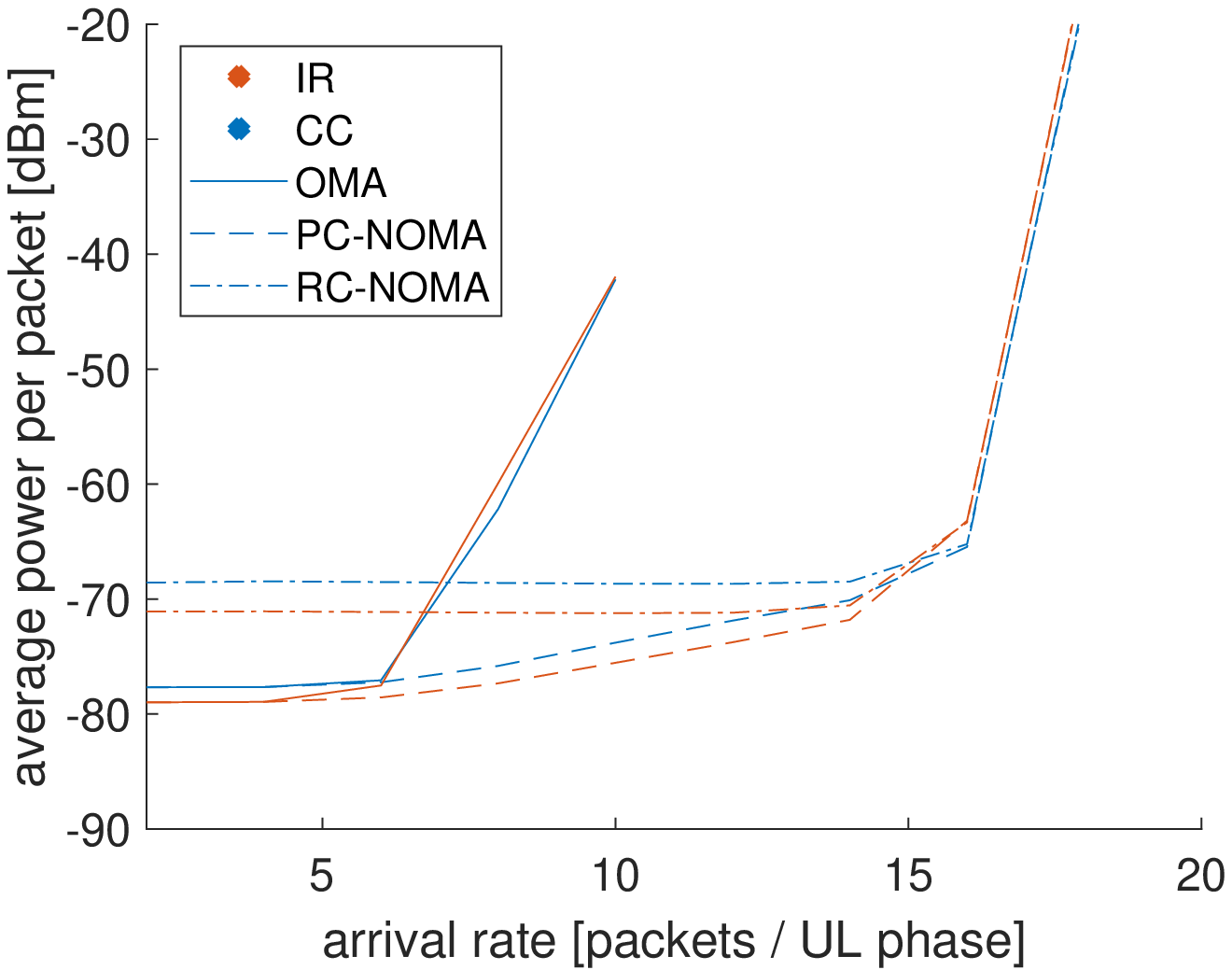}}%
\hfill
\subfigure[]{
\label{fig:ppp:finite}%
\includegraphics[width=0.33\textwidth]{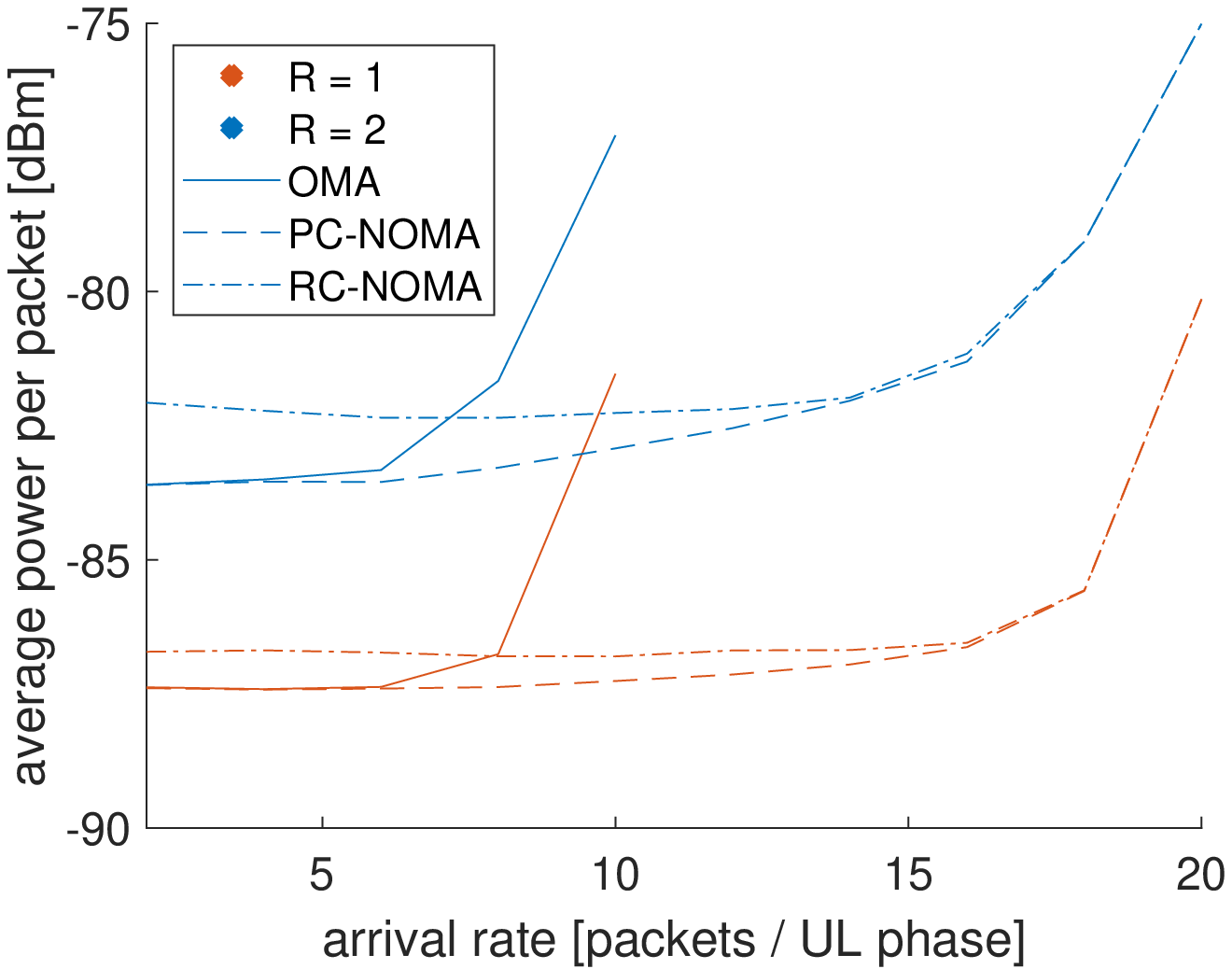}}%
\caption{Average power spent per packet. (a) Chase Combining (b) CC and IR at $R=2$ (c) Finite blocklength IR.}
\label{fig:res:2}
\end{figure}

In Fig.~\ref{fig:res:2} the average power spent per packet (i.e. including retransmissions) as a function of arrival rate is investigated in different configurations. 
Note that in these and other figures the results for OMA are only shown until $bN=10$, since at higher intensities the system is in a state of almost permanent availability outage.
In Fig.~\ref{fig:ppp:cc} the mode used is CC while the two sets of curves (red and blue) correspond to different transmission rates $R$. 
For very low arrival rates ($bN\in[2,4]$) OMA and PC-NOMA are equivalent. As the arrival rate increases, the PC-NOMA approach quickly becomes much more efficient than the baseline scheme. This leads to one of the main takeaways of this work: in a latency-constrained system with high reliability requirements, the largest power penalty comes from the necessity to queue the packets. 
While scheduling them in a non-orthogonal way introduces penalty of its own, it is in fact less detrimental
than having to make up for the lost transmission opportunities with more aggressive error targets.
By comparing the difference between PC-NOMA (dashed) and RC-NOMA (dot-dashed) we can see that this is especially the case for low transmission rates $R$ ($1$dB of difference between red set of curves and $9$dB for $R=2$).

In Fig.~\ref{fig:ppp:cc_vs_ir} CC (blue) and IR (red) at $R=2$ are compared. Application of the latter method allows to further improve the performance by lowering the average power by $~1.5$dB in case of OMA/PC-NOMA and $~2.5$dB with RC-NOMA. We note that towards higher arrival rates CC gains an upper hand over IR since its slightly lower initial error targets make it less likely to queue the packets.
Although the difference is minor, it reveals that obtaining a truly optimal solution would require adapting the error targets based on the current state of the buffer $T$ and knowledge of the arrival rate as well\footnote{However, as noted the room for improvement is not large and would add significant complexity to an already difficult problem. Last but not least, the information about the arrival rate in many scenarios might not be readily available.}.

Lastly, Fig.~\ref{fig:ppp:finite} depicts the results corresponding to the finite blocklength scenario with known channel. 
The availability of instantaneous CSI allows to greatly decrease the mean power compared to the statistical CSI case. In the low to moderate traffic range ($bN\leq 14$) savings reach $4.5$dB at $R=1$ and $11$dB at $R=2$. 
Furthermore, as the arrival rate grows the increase in required power is much slower in the Finite IR case than for the statistical CSI counterparts.

\begin{figure}[H]	
	\centering
	\includegraphics[width=0.5\linewidth]{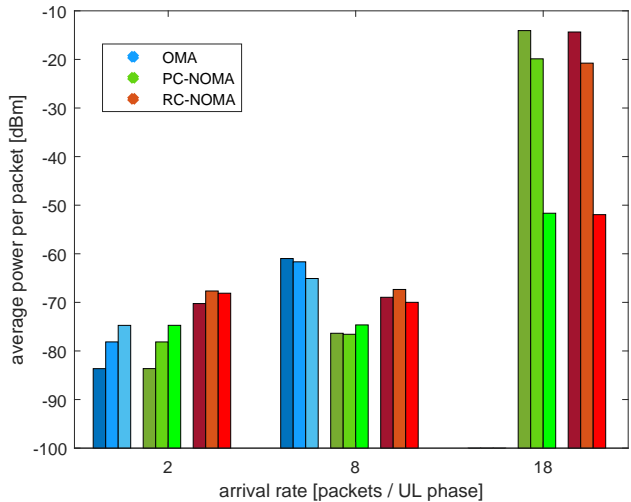}
	\caption{The average power per packet divided by zones. The three consecutive columns with different shades of the same color correspond to the average power per packet in a close (dark), middle and far (bright) zone around the BS. For the scenario considered here these are 20-53,33 meters, 53,33 - 86,66 meters, 86,66 - 120 meters.}
	\label{fig:res:3}
\end{figure}
In Fig.~\ref{fig:res:3} we investigate in more detail the average power per packet metric by looking at the performance of users grouped in different zones around the BS. As an example we take the Chase Combining case at $R=2$ and low, medium and high arrival rate ($bN=[2,8,18]$). 
Most notably, as the intensity of traffic increases, the burden is shifted to the users close to the BS. The reason is twofold. The first cause is again related to queuing which typically introduces lower penalty for UEs closer to the BS\footnote{As described in Section \ref{sec:scheduling:oma} the process of deciding which UEs to postpone is slightly more complex and ultimately depends also on the residual SNR/MI and remaining error target.
Nevertheless, packets from UEs which are positioned further away are less likely to be queued.}.
Another cause is specific to NOMA, which in order to work requires that one packet has higher received power than the other.
Since raising the power of UEs that are close is cheaper, typically they will be the ones asked to boost it (this behavior can be observed for RC-NOMA from the beginning).
Moreover, in a PC-NOMA at low to moderate arrival rates, only few pairs are needed so they are often created among UEs positioned closer to the BS, while the furthest users are assigned the remaining TF-blocks in an orthogonal manner.
Similar effects as those described have been observed also for lower transmission rates $R$ and in finite blocklength scenarios. 

\begin{figure}[t!]%
\centering
\subfigure[Chase combining]{
\label{fig:util:cc}%
\includegraphics[width=0.33\textwidth]{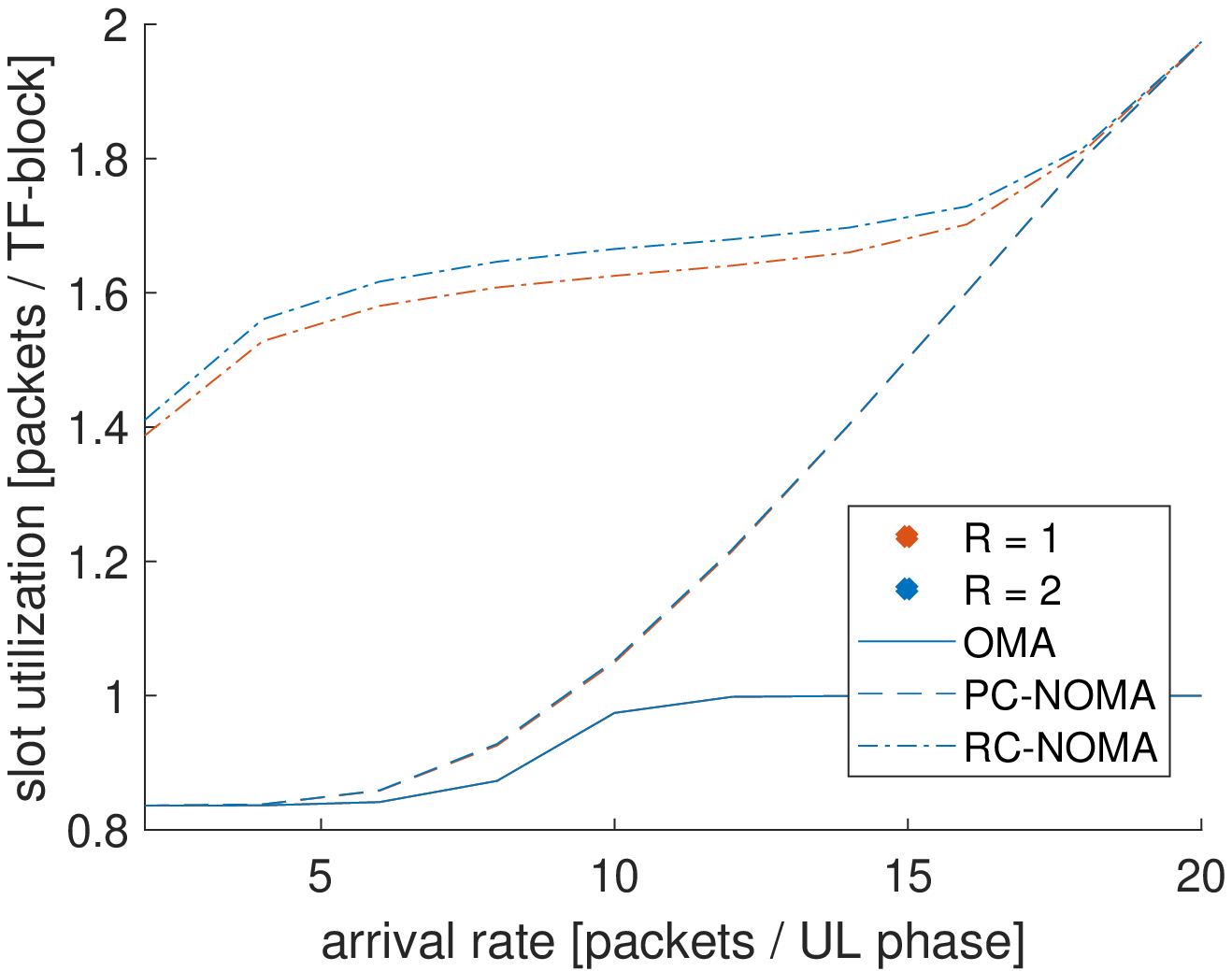}}%
\hfill
\subfigure[Incremental redundancy]{
\label{fig:util:ir}%
\includegraphics[width=0.33\textwidth]{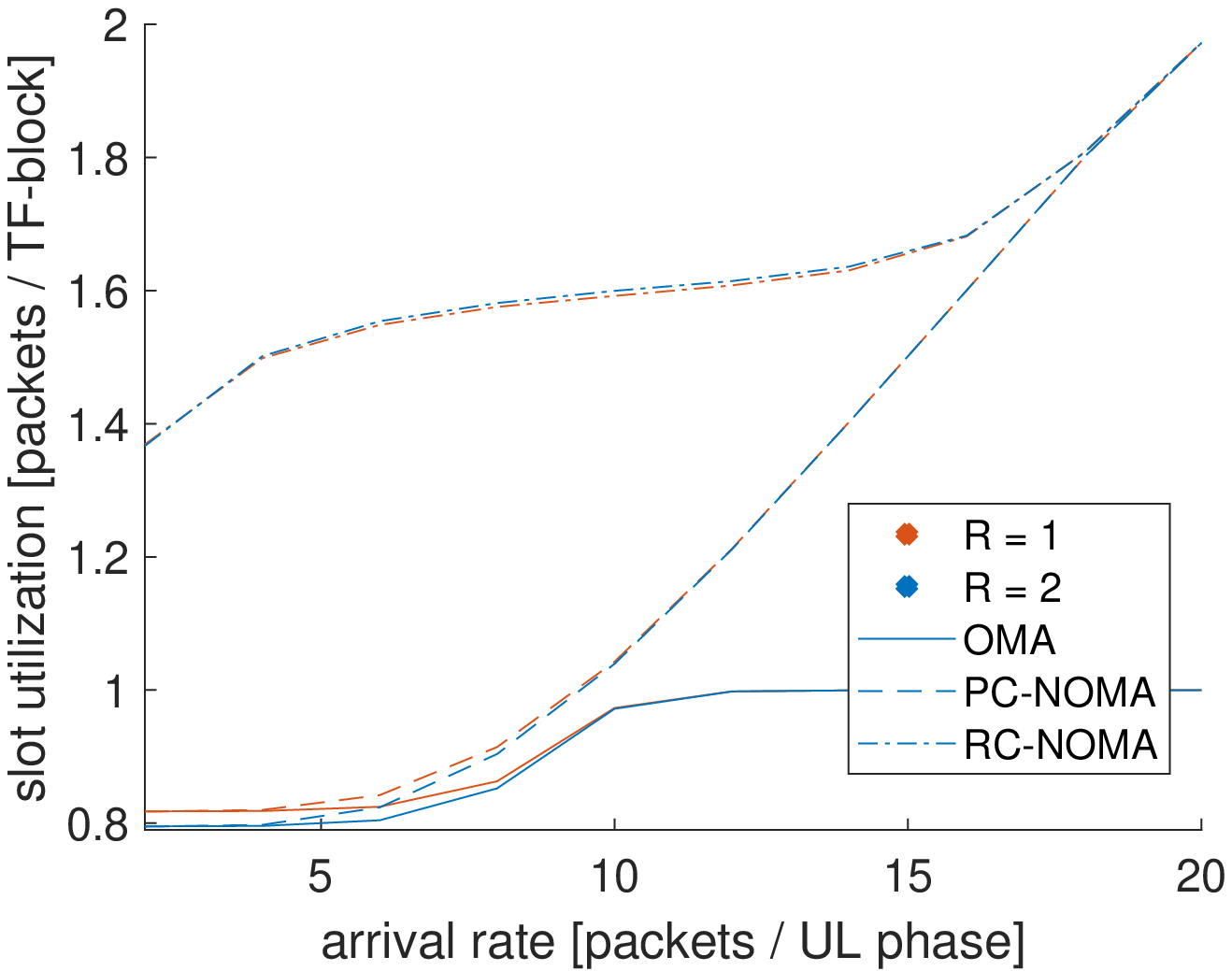}}%
\hfill
\subfigure[Finite IR]{
\label{fig:util:finite}%
\includegraphics[width=0.33\textwidth]{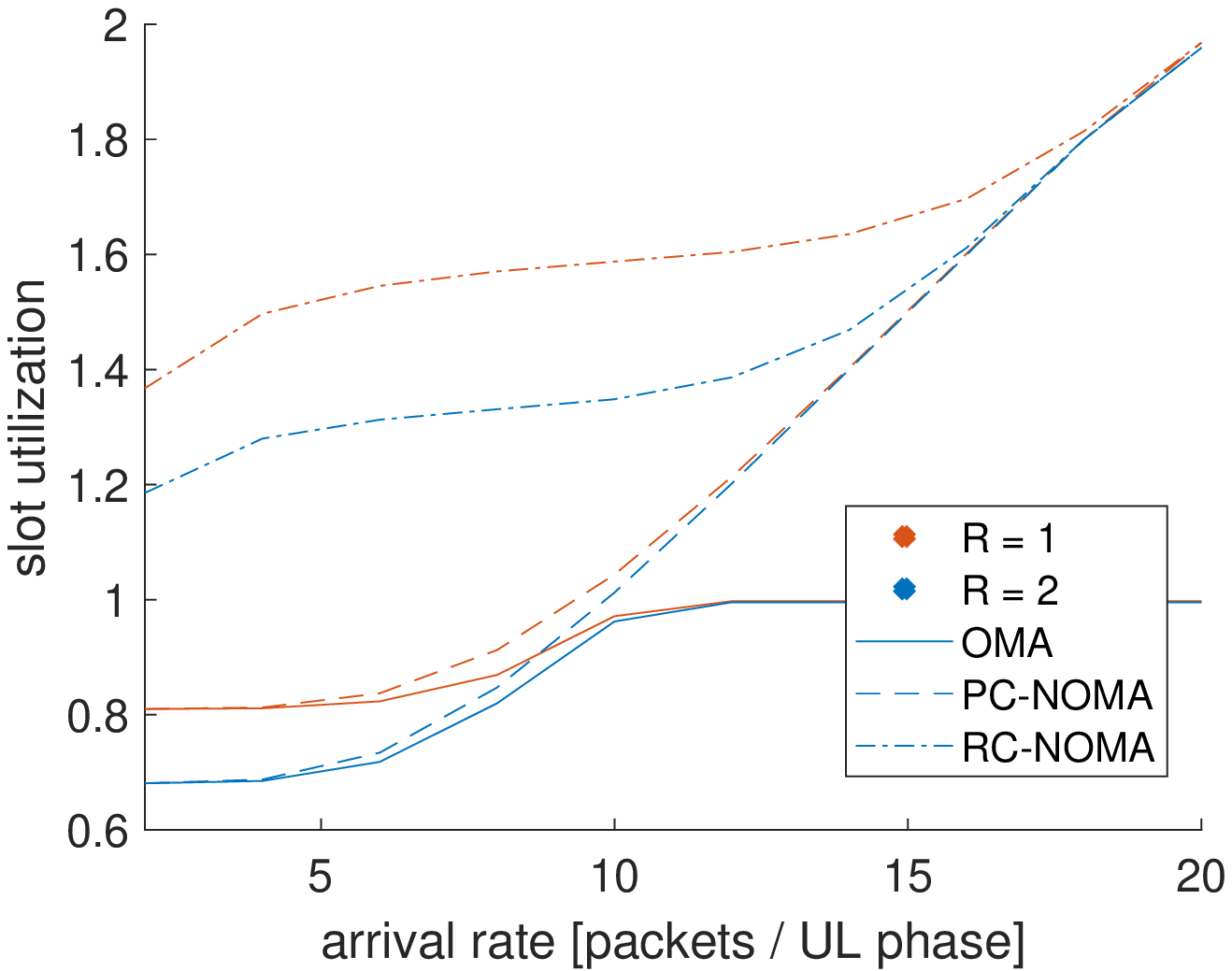}}%
\caption{Slot utilization of the studied access methods as a function of the arrival rate.}
\label{fig:res:4}
\end{figure}

Another set of results is provided in Fig.~\ref{fig:res:4}. We define the slot utilization as the total number of successfully decoded packets from all UEs divided by the total number of used TF-blocks. 
The dependency of slot utilization on retransmission mode IR/CC and rate follow the same discussion as earlier for Fig.~\ref{fig:res:1}. The higher the initial error targets, the more retransmissions are needed thus degrading the performance.
Between PC-NOMA and RC-NOMA, the more aggressive pairing strategy can clearly offer significant gains.
The reader is encouraged to analyze this especially in conjunction with Fig.\ref{fig:res:2}.
Observe that for low rate $R=1$ and low-to-medium traffic RC-NOMA almost doubles the resource efficiency of PC-NOMA with very little penalty to the average power (around $1$dB). For higher transmission rates the increase in average power is more significant so the choice between PC and RC variant becomes a matter of trade-off.

\begin{figure}[H]%
\centering
\subfigure[]{
\label{fig:rate:ppp}%
\includegraphics[width=0.42\textwidth]{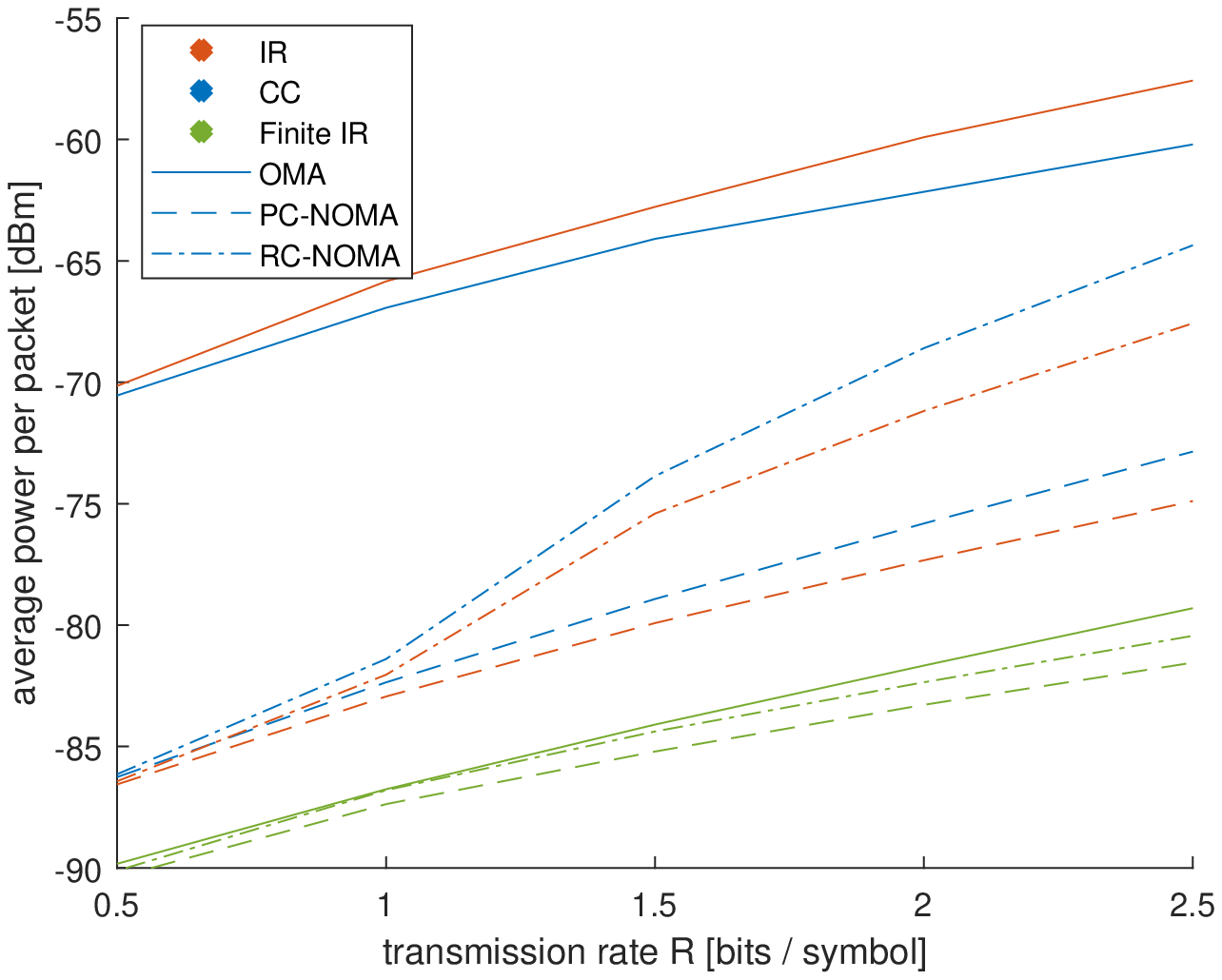}}%
\hfill
\subfigure[]{
\label{fig:rate:spectral}%
\includegraphics[width=0.42\textwidth]{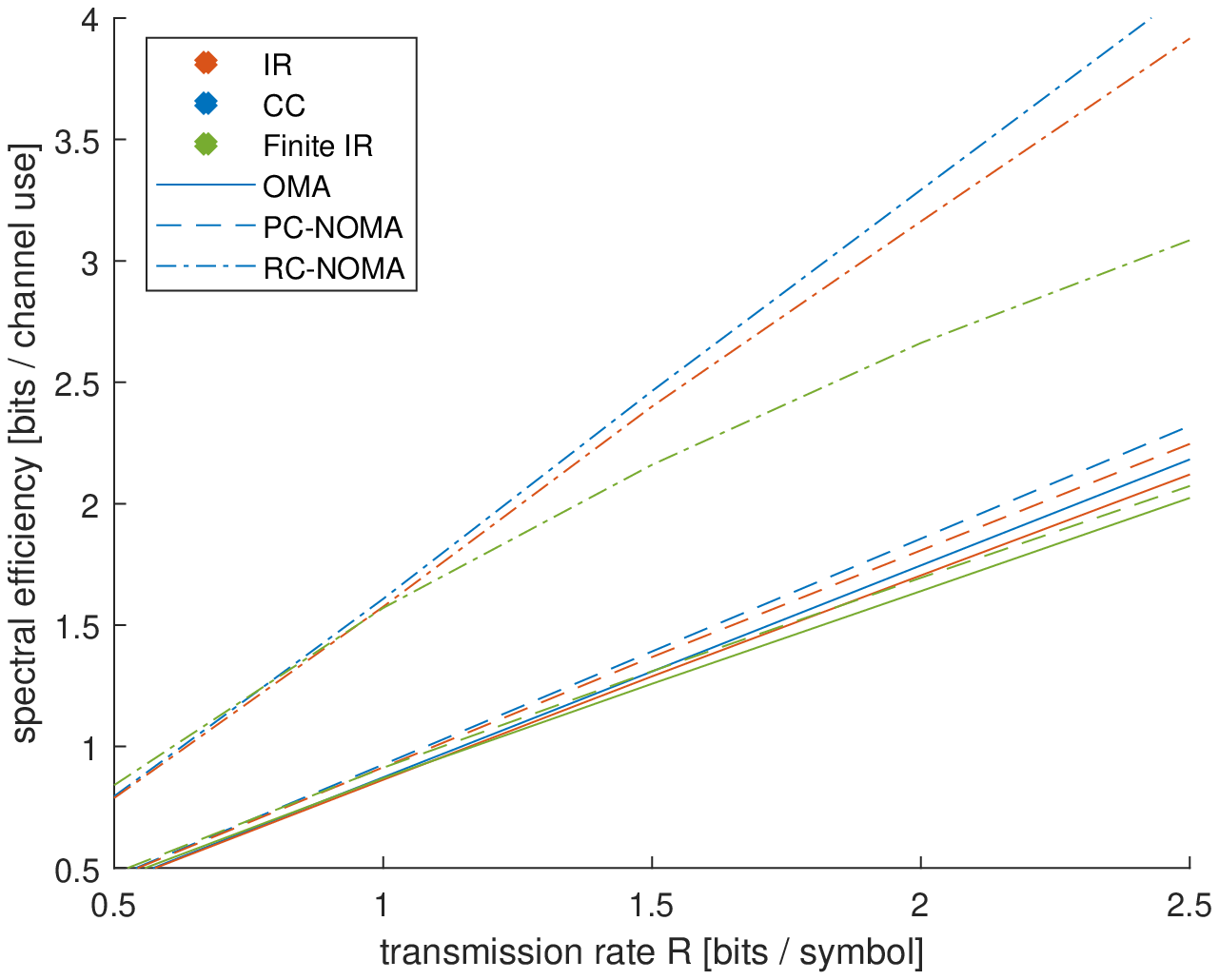}}%
\caption{
Average power per packet and spectral efficiency as a function of transmission rate at $bN=8$ [packets / UL phase].}
\label{fig:res:5}
\end{figure}

Lastly, in Fig.~\ref{fig:res:5} we fix the average arrival rate of new packets to $bN=8$ and instead vary the transmission rate $R$. The spectral efficiency presented in \ref{fig:rate:spectral} is obtained as the product of slot utilization and $R$.
The noticeable jump in power of RC-NOMA with statistical CSI above $R>1$ is in line with the observations first made in \cite{kotaba:icc}.
This behavior can be explained by inspecting the result \eqref{eq:err2closed}, which contains a special term that decreases the error probability whenever $\gamma_{j}^{(l)}\gamma_{k}^{(m)} \zeta_j \zeta_k < 1$. 
Since $\gamma_{j}^{(0)} = 2^R-1$ and $ a<b \iff \gamma_{j}^{(a)} \geq \gamma_{j}^{(b)}$, then the condition $\gamma_{j}^{(l)}\gamma_{k}^{(m)} \zeta_j \zeta_k < 1$ is always true for $R\leq 1$.
The similar jump in PC-NOMA is not observed at this arrival rate due to the fact that with $bN$ only equal to $8$, pairs are still relatively infrequent. Moreover, most of the time pairing between two new packets can be avoided. Instead, it is possible to transmit them on dedicated slots, while only the ones with $\gamma_j^{(l)},\gamma_k^{(m)}<2^R-1$ are combined so that $\gamma_j^{(l)}\gamma_k^{(m)}\zeta_j \zeta_k<1$.

\section{Conclusions}\label{sec:concl}
In this work we have proposed and investigated the performance of the system which combines NOMA and HARQ mechanisms to efficiently serve uplink URLLC traffic.
Two distinct scenarios were discussed: one where only statistical CSI is available, and another where additionally also the instantaneous channel realizations are known.
In each case we have defined an optimization problem that aims to minimize the average power spent per packet under a given latency (reflected by the maximum number of retransmissions) and reliability constraint.
The schemes were evaluated in a multi-user scenario with fixed amount of channel resources and varying traffic intensity to investigate the impact of queuing on the overall reliability, power and resource efficiency.
Our findings show that the introduction of NOMA is especially promising in two cases. 
First (RC-NOMA), the technique can be used to increase the total capacity of the system up to two times at a low-to-moderate cost in terms of power.
Second (PC-NOMA), it can be implemented as an emergency mechanism in situations where due to higher traffic demand using traditional OMA would lead to prohibitively high power or even complete availability outage.
The latter case is especially interesting as it shows that, in a latency-constrained system with given reliability requirements, the typical power penalty associated with NOMA is significantly smaller than the one arising from queuing the packets.
Lastly, by investigating each scheme in two CSI cases, we provide some insights into the bounds on achievable performance in practical scenarios.
Especially prominent is how the availability of instantaneous CSI can greatly reduce the transmit power needed for achieving the reliability targets.

\appendices
\section{Proof of Theorem \ref{theorem:1}}\label{sec:appendix:cc_proof}

\begin{proof}
The proof is split into two parts. The first claim is proven by induction as follows.\\

\subsubsection*{The induction step}
Assume that there exists a certain round $l$ where the optimal error $\epsilon_j^{(l)}$ minimizing the average power $\Psi_j^{(l)} (\gamma_j^{(l)},\Theta_j^{(l)})$ depends only on the remaining final error target $\Theta_j^{(l)}$, such that $\Psi_j^{(l)} (\gamma_j^{(l)},\Theta_j^{(l)}) = \gamma_j^{(l)}d_j^{\alpha}\sigma^2 \widetilde{\Psi}_j^{(l)} (\Theta_j^{(l)})$. 
If this is the case, then the optimization problem at an earlier round $l-1$ becomes
\begin{argmini}|s|[2]           	
    {\epsilon_j^{(l-1)}}           	
    {-\frac{\gamma_{j}^{(l-1)}d_j^{\alpha}\sigma^2}{\ln(1-\epsilon_j^{(l-1)})} + \int_{0}^{\gamma_{j}^{(l-1)}} f_e \left( x; \frac{P_j^{(l-1)}}{d_j^{\alpha}\sigma^2} \right) (\gamma_j^{(l-1)}-x)d_j^{\alpha}\sigma^2\widetilde{\Psi}_j^{(l)} \left(\frac{\Theta_j^{(l-1)}}{\epsilon_j^{(l-1)}} \right) dx}   
    {\label{eq:optProof}}   	
    {}                            
\end{argmini}
where the update $\gamma_j^{(l)} = \gamma_j^{(l-1)} - SNR_j^{(l-1)}$ is specific to CC and follows from \eqref{eq:errB}.
The objective function, which requires only simple integration can be obtained in the closed form
\begin{equation}
    -\gamma_j^{(l-1)}d_j^{\alpha}\sigma^2 \left( \frac{1}{\ln(1-\epsilon_j^{(l-1)})} - \frac{\ln(1-\epsilon_j^{(l-1)})+\epsilon_j^{(l-1)}}{\ln(1-\epsilon_j^{(l-1)})} \widetilde{\Psi}_j^{(l)} \left(\frac{\Theta_j^{(l-1)}}{\epsilon_j^{(l-1)}} \right) \right)
    \label{eq:induction_step}
\end{equation}
It is clear from the expression \eqref{eq:induction_step} which has a form $af(x)$, that the $\epsilon_j^{(l-1)}$ which minimizes it depends only on $\Theta_j^{(l-1)}$.

\subsubsection*{The basis step}
Since $\epsilon_j^{(L)}=\Theta_j^{(L)}=\frac{\epsilon_{tar}}{\prod_{i=0}^{L-1}\epsilon_j^{(i)}}$ used in the last possible transmission is fully determined by earlier attempts, the first non-trivial term corresponds to $\Psi_j^{(L-1)} (\gamma_j^{(L-1)},\Theta_j^{(L-1)})$. The objective function there, which we denote for short $P_j^{avg}$, reads
\begin{equation}
\begin{split}
&P_{j}^{avg} = P_j^{(L-1)} + \int_{0}^{\gamma_{j}^{(L-1)}}f_e\left( x_{L-1};\frac{P_j^{(L-1)}}{d_j^{\alpha}\sigma^2}\right) \left( -\frac{(\gamma_{j}^{(L-1)} -x_{L-1}) d_j^{\alpha}\sigma^2}{\ln(1-\epsilon_j^{(L)})} \right) dx_{L-1}\\
& = \left(-\gamma_{j}^{(L-1)}d_j^{\alpha}\sigma^2 \right) \left( \frac{1}{\ln(1-\epsilon_j^{(L-1)})} + \frac{\ln(1-\epsilon_j^{(L-1)})+\epsilon_j^{(L-1)}}{\ln(1-\epsilon_j^{(L)})\ln(1-\epsilon_j^{(L-1)})} \right)
\end{split}
\end{equation}
While solving $\frac{d P_{j}^{avg}}{d \epsilon_j^{(L-1)}}=0$ requires numerical method it is again clear that the result is independent of $\gamma_{j}^{(L-1)}$, $d_j^{\alpha}$ or $\sigma^2$.\\

Applying the induction to the basis step proves sequentially that in all rounds $L-1,\dots,1,0$ the optimal error target depends only on the current error budget.
As for the second claim of the theorem,
notice that when the optimal error targets do not depend on the residual SNRs, it means that for each round $l$ they must have a single, well-defined value, which can be computed in advance.
This is because fixing $\epsilon_j^{(l)}$ 
leads to a chain of uniquely determined values $\epsilon_j^{(l)} \rightarrow \frac{\Theta_j^{(l)}}{\epsilon_j^{(l)}} \xrightarrow{opt} \epsilon_j^{(l+1)} \rightarrow \frac{\Theta_j^{(l)}}{\epsilon_j^{(l)}\epsilon_j^{(l+1)}} \xrightarrow{opt} \dots \xrightarrow{opt} \epsilon_j^{(L)}$.
By writing the problem \eqref{eq:optGlob} in its explicit form 
and using the fact that error targets do not depend on the residual SNRs and hence on the variables of integration it is possible to eventually arrive at \eqref{eq:optSimple}.
The derivation is relatively simple albeit quite tedious. Although calculations involve multiple nested integrals, all integrands are of the form either $ae^x$ or $axe^x$ and display a regular structure.

\end{proof}

\section{}\label{sec:appendix:closed_form}
Here, we will show the derivation of \eqref{eq:err2closed} from \eqref{eq:err2U}.
First, let us shorten the notation by introducing following quantities: $X \sim f_e\left( x;s \right)$ where $s = \frac{P_j^{(l)}}{d_j^{\alpha}}$ is the exponentially distributed received power from user $j$ and similarly $Y \sim f_e\left(y;p \right)$ where $p = \frac{P_k^{(m)}}{d_k^{\alpha}}$ corresponds to user $k$. 
Also, since only a single packet from each user is considered we can drop the superscripts $(l)$ and $(m)$ moving forward.
The first probability component in \eqref{eq:err2U} now reads:
\begin{equation}
    \mathrm{Pr}\left\lbrace \frac{X}{\sigma^2} < \gamma_j , \frac{Y}{X \zeta_j + \sigma^2} > \gamma_k\right\rbrace = \int_{0}^{\gamma_j \sigma^2} \left( \int_{\gamma_k \left( x\zeta_j+\sigma^2\right)}^{\infty} \frac{1}{p}e^{-\frac{y}{p}} dy \right) \frac{1}{s} e^{-\frac{x}{s}} dx
    \label{eq:app1}
\end{equation}
while the second term
\begin{equation}
    \mathrm{Pr}\left\lbrace \frac{X}{Y\zeta_k+\sigma^2} < \gamma_j , \frac{Y}{X \zeta_j + \sigma^2} < \gamma_k\right\rbrace = \int_{0}^{\infty} \left( \int_{\frac{x}{\gamma_j\zeta_k}-\frac{\sigma^2}{\zeta_k}}^{\gamma_k \left( x\zeta_j+\sigma^2\right)} \frac{1}{p}e^{-\frac{y}{p}} dy \right) \frac{1}{s} e^{-\frac{x}{s}} dx
    \label{eq:app2}
\end{equation}
Notice that when $x < \gamma_j\sigma^2$, the lower limit of the inner integral in \eqref{eq:app2} is negative and outside of the support of the exponential distribution. Hence we can write \eqref{eq:app2} instead as:
\begin{equation}
    \int_{0}^{\gamma_j\sigma^2} \left( \int_{0}^{\gamma_k \left( x\zeta_j+\sigma^2\right)} \frac{1}{p}e^{-\frac{y}{p}} dy \right) \frac{1}{s} e^{-\frac{x}{s}} dx + 
    \int_{\gamma_j\sigma^2}^{\infty} \left( \int_{\frac{x}{\gamma_j\zeta_k}-\frac{\sigma^2}{\zeta_k}}^{\gamma_k \left( x\zeta_j+\sigma^2\right)} \frac{1}{p}e^{-\frac{y}{p}} dy \right) \frac{1}{s} e^{-\frac{x}{s}} dx
    \label{eq:app3}
\end{equation}
The expression \eqref{eq:app1} and the first term in \eqref{eq:app3} complement each other so their sum becomes
\begin{equation}
    \int_{0}^{\gamma_j \sigma^2} \left( \int_{0}^{\infty} \frac{1}{p}e^{-\frac{y}{p}} dy \right) \frac{1}{s} e^{-\frac{x}{s}} dx = 
    \int_{0}^{\gamma_j \sigma^2} \frac{1}{s} e^{-\frac{x}{s}} dx = F_e \left(\gamma_j \sigma^2 ; s \right)
    \label{eq:app4}
\end{equation}
The second component of \eqref{eq:app3} is slightly more involved. First, let us focus on the relationship between the limits of its second integral. After rearranging the terms we obtain:
\begin{equation}
    x \left( \gamma_k \zeta_j - \frac{1}{\gamma_j \zeta_k} \right) \geq -\gamma_k \sigma^2 - \frac{\sigma^2}{\zeta_k}.
    \label{eq:app6}
\end{equation}
Since the right side is negative and $x > 0$, then \eqref{eq:app6} is always true whenever $ \gamma_k \zeta_j - \frac{1}{\gamma_j \zeta_k} > 0$ leading to no additional constraint on $x$. However, when $\gamma_k \zeta_j - \frac{1}{\gamma_j \zeta_k}$ is negative, or equivalently $\gamma_j\gamma_k\zeta_j\zeta_k < 1$, then the upper limit on $x$ appears:
\begin{equation}
    x \leq \frac{\gamma_j\gamma_k\zeta_k \sigma^2 + \gamma_j \sigma^2}{1 - \gamma_j\gamma_k\zeta_j\zeta_k}
\end{equation}
which is a valid limit since $\frac{\gamma_j\gamma_k\zeta_k \sigma^2 + \gamma_j \sigma^2}{1 - \gamma_j\gamma_k\zeta_j\zeta_k} > \frac{\gamma_j \sigma^2}{1 - \gamma_j\gamma_k\zeta_j\zeta_k} > \gamma_j \sigma^2$. 
The missing integral yields
\begin{equation}
\begin{aligned}
    & \int_{\gamma_j\sigma^2}^{C} \left( \int_{\frac{x}{\gamma_j\zeta_k}-\frac{\sigma^2}{\zeta_k}}^{\gamma_k \left( x\zeta_j+\sigma^2\right)} \frac{1}{p}e^{-\frac{y}{p}} dy \right) \frac{1}{s} e^{-\frac{x}{s}} dx 
    = \frac{1}{s}  \int_{\gamma_j \sigma^2}^{C} e^{\frac{\sigma^2}{\zeta_k p}} e^{-x\frac{s+\gamma_j \zeta_k p}{\gamma_j \zeta_k p s}} - e^{-\frac{\gamma_k \sigma^2}{p}} e^{-x\frac{\gamma_k\zeta_j s + p}{ps}} dx\\
    &= \frac{\gamma_j \zeta_k p}{s+\gamma_j \zeta_k p}\left( e^{-\frac{\gamma_j \sigma^2}{s}} - e^{-C \frac{s+\gamma_j \zeta_k p}{\gamma_j \zeta_k p s} + \frac{\sigma^2}{\zeta_k p}}\right) - \frac{p}{p+\gamma_k\zeta_j s}\left( e^{-\frac{\gamma_j \sigma^2}{s}} e^{-\gamma_k \sigma^2 \frac{1+\gamma_j \zeta_j}{p}} - e^{-C\frac{\gamma_k \zeta_j s + p}{p s} - \frac{\gamma_k\sigma^2}{p}} \right).
    \label{eq:appF}
\end{aligned}
\end{equation}
When $\gamma_j\gamma_k\zeta_j\zeta_k > 1$ the second and fourth term in \eqref{eq:appF} disappear since $\displaystyle{\lim_{C\to \infty}} e^{-C \frac{s+\gamma_j \zeta_k p}{\gamma_j \zeta_k p s} + \frac{\sigma^2}{\zeta_k p}} = 0$ and $\displaystyle{\lim_{C\to \infty}} e^{-C\frac{\gamma_k \zeta_j s + p}{p s} - \frac{\gamma_k\sigma^2}{p}} = 0$. Otherwise, $ C = \frac{\gamma_j\gamma_k\zeta_k \sigma^2 + \gamma_j \sigma^2}{1 - \gamma_j\gamma_k\zeta_j\zeta_k}$ and after some simplification we obtain that $e^{-C \frac{s+\gamma_j \zeta_k p}{\gamma_j \zeta_k p s} + \frac{\sigma^2}{\zeta_k p}} = e^{-C\frac{\gamma_k \zeta_j s + p}{p s} - \frac{\gamma_k\sigma^2}{p}} = e^{-\frac{\sigma^2}{1-\gamma_j \gamma_k \zeta_j \zeta_k} \left(\gamma_j \frac{\gamma_k \zeta_k + 1}{s} + \gamma_k \frac{\gamma_j \zeta_j +1}{p} \right)}$. The total error probability is then the sum of \eqref{eq:app4} and \eqref{eq:appF}.

\section{}\label{sec:appendix:reduction}

Let us consider a received signal over a single TF-block given by
\begin{equation}
\*y' = h_1^{(a)}\*x_1 + h_2^{(b)}\*x_2 + \*n_1 
\label{eq:reduction1}
\end{equation}
and let us assume that in one of the previous uplink phases the interferer (UE 2) already had an unsuccessful transmission attempt of the packet so the BS has stored
\begin{equation}
\*y'' = h_2^{(b-1)}\*x_2 + h_3^{(c)}\*x_3 + \*n_2
\label{eq:reduction2}
\end{equation}
where $h_1^{(a)}$, $h_2^{(b)}$, $h_2^{(b-1)}$ and $h_3^{(c)}$ denote the complex channel coefficients and the transmit power and path loss coefficients of each user were omitted for simplicity.
Instead of attempting to decode $\*x_1$ directly from $\*y'$ which would yield SINR equal to $\frac{\left| h_1^{(a)}\right|^2}{\left| h_2^{(b)}\right|^2+ \sigma^2}$ the receiver can consider signal $\*y' - q\*y''$ which yields SINR $\frac{\left| h_1^{(a)}\right|^2}{\left| h_2^{(b)}-q h_2^{(b-1)}\right|^2 + \left| q\right|^2 \left( \left| h_3^{(c)}\right|^2+ \sigma^2 \right) + \sigma^2}$.
The expression is maximized for $q=\frac{h_2^{(b)} h_2^{(b-1)^\ast}}{\left| h_2^{(b-1)}\right|^2+\left| h_3^{(c)}\right|^2+ \sigma^2}$ in which case the the SINR becomes
$\frac{\left| h_1^{(a)}\right|^2}{\left| h_2^{(b)}\right|^2\frac{\left| h_3^{(c)}\right|^2+ \sigma^2}{\left| h_2^{(b-1)}\right|^2 + \left| h_3^{(c)}\right|^2+ \sigma^2} + \sigma^2}.
\label{eq:snr_scale}$
It's easy to notice that, compared to \eqref{eq:reduction1}, the power of the interfering component $\left| h_2^{(b)}\right|^2$ is now scaled down by a factor
\begin{equation}
    \zeta_2
    =\frac{\left| h_3^{(c)}\right|^2+ \sigma^2}{\left| h_2^{(b-1)}\right|^2 + \left| h_3^{(c)}\right|^2+ \sigma^2}
    =\left(1+\frac{\left| h_2^{(b-1)}\right|^2}{\left| h_3^{(c)}\right|^2+ \sigma^2} \right)^{-1}.
\end{equation}
The amount is directly related to the SINR that UE 2 experienced in its past replica \eqref{eq:reduction2}.

Note that the operation described above has this particularly simple form only when 
the signal $\*y''$ used to reduce the interference is uncorrelated with the symbols $\*x_1$, but this is ensured already since in CC we do not allow $\*x_1$ and $\*x_2$ to be paired together twice.

\ifCLASSOPTIONcaptionsoff
  \newpage
\fi

\end{document}